\documentclass[preprint,aps,superscriptaddress,showkeys,11pt]{revtex4}
\pdfoutput=1
\usepackage{graphicx}
\usepackage{amsmath}
\usepackage{amsfonts}
\usepackage{amssymb}
\usepackage{xcolor, soul}
\usepackage{epstopdf}
\usepackage{float}
\usepackage{subfigure}
\usepackage{hyperref}
\hypersetup{
     colorlinks   = true,
     citecolor    = red,
     linkcolor    = blue,
     urlcolor     = blue,
}
\usepackage{cleveref}

\def\beq{\begin{equation}}
\def\eeq{\end{equation}}
\def\bea{\begin{eqnarray}}
\def\eea{\end{eqnarray}}
\def\be{\begin{equation}}
\def\ee{\end{equation}}

\def\nno{\nonumber}
\def\bse{\begin{subequations}}
\def\ese{\end{subequations}}

\def\wI{w_{\phi}}
\def\phiast{\phi_{\ast}}

\graphicspath{{./figs/}}

\begin{document}

\title{CMB constraints on dark matter phenomenology via reheating in Minimal plateau inflation}

\author{Debaprasad Maity}
 \email{debu@iitg.ac.in}
\author{Pankaj Saha}%
 \email{pankaj.saha@iitg.ac.in}
\affiliation{%
Department of Physics, Indian Institute of Technology Guwahati.\\
 Guwahati, Assam, India 
}%

\date{\today}

\begin{abstract}
{We consider the CMB constrains on reheating and dark matter parameter space for a specific plateau type inflationary model. The plateau inflationary models which are currently most favored models from data can be well approximated by a potential of the form $V(\phi)\propto \phi^n$ around $\phi=0$. This fact makes it possible to study the reheating phase with general inflaton equation of state in a viable cosmological scenario. In addition, following our recent work\cite{Maity:2018dgy}, we generalize the connection between reheating and the present CMB data and the dark matter parameter space for general inflaton equation of state parameter.}
\end{abstract}
\maketitle
\tableofcontents
\newpage
\section{\label{intro}Introduction}
The idea of inflation\cite{guth,Starobinsky:1980te,Sato:1980yn,linde1,steinhardt} which was proposed around 1980's is currently the leading paradigm for a solution to the so-called \textit{initial conditions} problems of the hot big bang theory. The nearly scale-invariant power spectra and the observed inhomogeneities in the CMB can be remarkably explained by an inflationary era prior to the radiation dominated hot big bang evolution. However, the inflation provides a universe which is devoid of any matter or radiation and as a natural consequence the idea of a phase of reheating\cite{Albrecht:1982mp,Dolgov:1982th,Abbott:1982hn,Traschen:1990sw} was proposed shortly after the proposal of inflationary universe. During the reheating phase the inflaton decays to produce all the matter and radiation that constitute our universe. Thus the phase of reheating is extremely important from both theoretical and observational point of view for a complete understanding of the evolution and the present constituents of our universe. Although the inflationary phase has several constraints from the CMB data, the reheating phase is still poorly constrained due to lack of direct observables. From the theoretical point of view, understanding this phase is proven to be difficult as the phase comprises several highly non-linear stages. On the other hand various observations\cite{Rubin:1967msa,Rubin:1970zza,Hu:2001bc,Bertone:2004pz} show that the majority of matter in our present universe is in the form of dark matter. The dark matter has also shaped the observed patterns in the CMB. Despite significant efforts, the direct detection of dark matter is still not been possible. Till date the only known information about the dark matter is its relic abundance measured precisely by PLANCK\cite{Ade:2015xua} to be $\Omega_X h^2 = 0.1188\pm0.0010$. Hence, we could seek to constrain the dark matter parameter space from indirect methods. In view of these questions, we have proposed a formalism to connect the CMB parameter space with dark matter parameter in \cite{Maity:2018dgy}. We extended the earlier works on constraining reheating epoch from CMB data\cite{Dai:2014jja,Creminelli:2014fca,Martin:2014nya,Domcke:2015iaa} by incorporating the explicit inflaton decay. The inflationary models that we analyzed previously were all of the form of $V(\phi) \propto \phi^2$ during reheating. For the reheating phase, apart from the inflaton decay term, the most important factor is the equation of the state parameter of the inflaton. It has been pointed out that for inflaton oscillating in a potential $V(\phi) \propto \phi^n$, the equation of the state can be well approximated by using the virial theorem as\cite{Mukhanov:2005sc}
\begin{equation}
 w_{eff} = \frac{p_{\phi}}{\rho_{\phi}} \simeq \frac{\langle \phi V'(\phi) -2V(\phi)\rangle}{\langle \phi V'(\phi) + 2V(\phi)\rangle} = \frac{n-2}{n+2},
 \label{eos}
\end{equation}
which is simply given by a dust-like equation of the state parameter viz. $w_{eff}=0$ in a $\phi^2$ potential. It would be natural to inquire about the effects of different equation of the state parameters on the reheating phase. However, the observations has strictly restricted chaotic inflationary models with $n>2$. Interestingly, a class of inflationary models, known as the $\alpha$-attractors, gained a lot of attention recently. Apart from the power of unifying a large class of inflationary models into a single fold, a useful feature of these models is that the inflaton potential can be well approximated as $V(\phi)\propto\phi^n$ around $\phi=0$. These fact provide us with an opportunity to study the effects of various inflaton equation of state in a viable cosmological scenario. The main motivation of the present work concerns with studying those effects on reheating as well as to constrain the dark matter phenomenology with constrains from CMB and current dark matter relic abundance. The inflationary models that we will study in the present work is the recently proposed minimal plateau inflationary models\cite{Maity:2019ltu}. The potentials has similar properties of that of the $\alpha$-attractor models. Hence, the qualitative features that will found in the present work will be applicable to any inflationary models whose potential can be approximated as $\phi^n$ behavior during reheating. 

We organize our paper as follows: In section-\ref{model}, we briefly describe the minimal plateau inflationary models and its parameters. After the end of inflation, the inflaton, in general, starts to have coherent oscillation at the minimum of the potential, during which the universe undergoes reheating phase. We also compute the effective equation of state of the oscillating inflaton for our subsequent studies. In section-\ref{reheatingprediction}, we have studied the reheating constraints from CMB and production of a heavy dark matter particle during reheating. Finally, we concluded and discussed our future work.

\section{\label{model}A Brief account on minimal plateau inflationary potentials}
In this section we will briefly describe the inflationary models proposed in \cite{Maity:2019ltu}. The model is a non-polynomial modification to the usual power-law chaotic potentials $\phi^n$ and is given by,
\bea
V_{min}(\phi) =\lambda \frac{m^{4-n} \phi^n}{1+\left(\frac{\phi}{\phi_*}\right)^n} 
\eea
The index $n$ takes only even values as in the case of chaotic models. The parameter $\lambda$(~when $n=4$) or the scale $m$ has similar properties that of the chaotic models and their values are fixed from WMAP normalization\cite{Komatsu2010}. The new scale $\phiast$ in these class models controls the shape of the potentials and provides the extended plateaus in the field space of the potentials for large field values while retaining the potential minimum around $\phi=0$. The origin of the potentials either from a general scalar tensor theory with suitable conformal transformation to Einstein frame or from supergravity theory with an anomalous $U(1)$ symmetry has been described in the previous work. It has also been shown that the models matches extremely well with the latest PLANCK data\cite{Ade:2015lrj,Akrami:2018odb} for a wide range of the controlling scale $\phiast$. The non-perturbative particle production for the models has been studied in \cite{Maity:2018qhi}. The main motivation of this work, as mentioned above, is to study the CMB constraints on reheating and consequently on the dark matter parameter space following the formalisn proposed in \cite{ Maity:2018dgy}.
We will show how the inflationary scale $\phiast$ plays important role in determining the reheating temperature and consequently its influence on the dark matter parameter space. The Only assumption we will make in our analysis is that of the perturbative decay of the inflaton field into radiation while the dark matter is produced due to the radiation annihilation.  
 
\section{CMB to dark matter phenomenology via reheating}
\label{reheatingprediction}
It has been a general consensus that the inflation phase must be followed by the reheating phase. During this phase, the inflaton field is assumed to decay into various daughter fields such as radiation, dark matter particles, etc. The quantity which characterizes this phase is known as the reheating temperature $(T_{re})$ and its duration $(N_{re})$. Reheating temperature is defined at the instant when the inflaton decay rate equals the universe expansion rate and the radiation domination sets in.
 Being not an observable quantity, the reheating temperature is usually considered as a free parameter where its upper value could be as high as the GUT scale and the lower value is fixed by the requirement of
  successful Big Bang Nucleosynthesis. However, it has been established in the recent reference \cite{Dai:2014jja} that not just inflationary phase but also  the reheating phase can be constrained by CMB anisotropy. Since then, there has been a surge of research work in this direction\cite{Creminelli:2014fca,Martin:2014nya,Cook:2015vqa,Ellis:2015pla,Ueno:2016dim,Eshaghi:2016kne,DiMarco:2017zek,Bhattacharya:2017ysa,Drewes:2017fmn}. The inherent assumption in all those works is that the inflaton field converts into radiation instantaneously at the end of reheating and the effective fluid describing the reheating dynamics is parameterized by an average equation of state parameter. In our recent work\cite{Maity:2018dgy,mhiggs}, we have shown that by relaxing the aforementioned assumptions we not only can constrain the inflationary model but also shed light on the dark matter
   phenomenology. Through our analysis, we were able to connect the current dark matter relic abundance with the temperature anisotropy in CMB. In this paper, we will follow the aforementioned formalism developed in \cite{Maity:2018dgy} and generalize it for the arbitrary equation of state of the inflaton during inflation. In our model, during reheating phase the inflaton potential can be approximated as $V(\phi) \propto \phi^n$ such that the equation of state of the
    \begin{figure}[t]
    	\centering	
    	\includegraphics[scale=0.8]{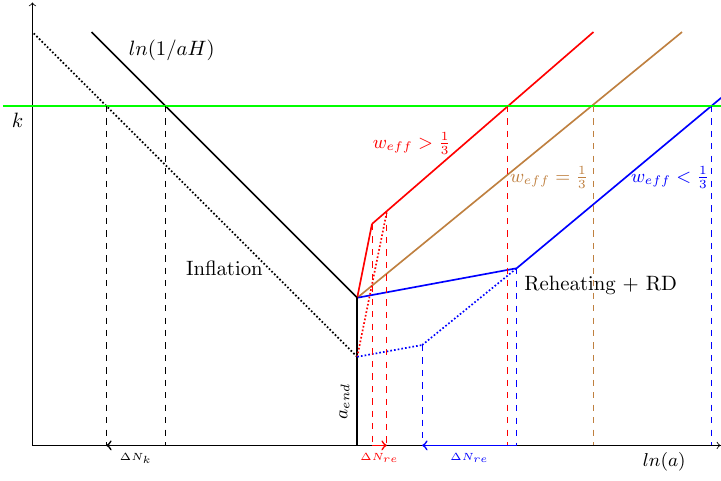}
    	\caption{\scriptsize Evolution of the comoving Hubble horizon starting from inflation (left side black lines), reheating (intermediate lines), radiation domination up to the present time(the shorter period of matter domination and the current dark energy domination is not shown as they don't affect the qualitative conclusion). During the reheating phase when the effective equation of state $w_{eff} = 1/3$, the evolution is indistinguishable from radiation dominated phase as shown in solid brown line. This fact makes reheating parameters $(N_{re}, T_{re})$ indeterministic as observed in \cite{Dai:2014jja}. The evolution with $w_{eff}<1/3$ is shown in blue intermediate lines while that of $w_{eff}> 1/3$ is shown in red intermediate lines. Increase in $n_s$ implies going from solid black line to dotted black line adding extra e-folding number $\Delta N_k$ for horizon crossing of a particular CMB mode $k$. It is clear from the figure for $w_{eff}<1/3$, increasing $n_s$ will result in decreasing reheating e-folding number during reheating $N_{re}$ thereby increasing reheating temperature $T_{re}$. While for $w_{\phi}>1/3$ it will just be the opposite. Hence, one can clearly expect an upper bound on $n_s$ for a particular inflation model with $w_{\phi}<1/3$ where the intermediate line closes indicating an instantaneous reheating with maximum possible reheating temperature. On the other hand for models with $w_{\phi}>1/3$, one expects a lower bound on $n_s$. The bound on the other end of $n_s$ for both the cases will be from the condition of minimum reheating temperature for a successful BBN.}  
    	\label{scales}
    \end{figure}
    oscillating inflaton can be approximated as $w_{\phi} = (n-2)/(n+2)$. An illustration of how CMB constraints the reheating temperature as well as how the equation of state parameter play a significant role controlling the reheating period is presented in fig.(\ref{scales}).
    It is also clear from the figure that for a particular scale and the inflaton equation of state during reheating we have maximum possible value of $n^{max}_s$ for $w_{eff} <1/3$ and minimum possible value of $n^{min}_s$ for $w_{eff} > 1/3$. Where $w_{eff}$ is the effective equation of state during reheating. However, for our case, we will consider explicit decay of inflation to radiation and dark matter during reheating. Therefore, the existence of $(n_s^{max},n_s^{min})$ will depend on the equation of state of inflaton $w_{\phi}$ itself. It also implies the existence of maximum reheating temperature $T_{re}^{max}$. Interestingly we will see that the maximum reheating temperature $T_{re}^{max}$ does not depend on a particular model under consideration.

\subsubsection{The Boltzmann Equations during reheating with general inflaton equation of state}
 We will consider three component universe consisting of inflaton($\rho_{\phi}$), radiation($\rho_R$) and a massive dark matter particle with the number density $(n_X)$. For simplicity, we consider dark matter particle production only from radiation annihilation. The thermal-averaged cross section
times velocity for the process is $\langle\sigma v\rangle$. For the general equation of state of inflaton, we will work with the following rescaled variables,
 \begin{eqnarray}
  \Phi = \frac{\rho_{\phi}a^{3(1+w_{\phi})}}{m_{\phi}^{(1-3w_{\phi})}};~~~~~~~~~~R = \rho_R a^4;~~~~~~~~~~X = n_X a^3. 
 \end{eqnarray}
The Boltzmann equations\cite{Giudice:2000ex,Maity:2018dgy} in the rescaled variables with the time variable $A=a/a_I$ reduces to, 
\begin{eqnarray}
\label{Inflatoneq}
\frac{d\Phi}{dA} &=& -c_1(1+ w_{\phi}) \frac{A^{1/2} \Phi}{\mathbb{H}};\\
\label{Radeq}
\frac{dR}{dA} &=& c_1(1+ w_{\phi}) \frac{A^{3(1-2 w_{\phi})/2}}{\mathbb{H}} \Phi + c_2 \frac{A^{-3/2} 2\left<E_X\right> \left<\sigma v\right> M_{pl}}{\mathbb{H}}\left(X^2 - X_{eq}^2\right);\\
\label{Xeq}
\frac{dX}{dA} &=& - c_2 \frac{A^{-5/2} 2\left<E_X\right> \left<\sigma v\right> M_{pl}}{\mathbb{H}}\left(X^2 - X_{eq}^2\right)\nno;
\end{eqnarray}
Where, $\mathbb{H} = \left(\Phi/A^{3 w_{\phi}} + R/A + X \left<E_X\right>/m_{\phi}\right)^{1/2}$ is the Hubble expansion rate and the constants $c_1$ and $c_2$ are given by
\begin{eqnarray}
c_1 = \sqrt{3}\frac{M_P \Gamma}{m_{\phi}^2},~~~~~c_2 = \sqrt{\frac{3}{8\pi}}
\end{eqnarray}
Here, $M_{pl}(=\sqrt{8\pi}M_p)$ is the Planck mass. $\left<E_X\right> = \rho_X/n_X$ is the average energy density of the X-particle which we identify as the candidate dark matter. $X_{eq}$ is the equilibrium (comoving-)number density of the dark matter of mass $M_X$ if it is in equilibrium with background thermal bath at temperature $T$. $\Gamma$ is the inflaton decay rate. We assume the dark matter has an internal degree of freedom $g$. Since we want to study all range of mass of the dark matter, we use the use the relativistic form of the equilibrium distribution function with the modified Bessel function of the second kind as\cite{Giudice:2000ex,book22}
\begin{eqnarray}
X_{eq} & = & \frac{g T^3}{2\pi^2} \left(\frac{M_X}{T}\right)^2K_2\left(\frac{M_X}{T}\right)   \left(\frac{A}{m_{\phi}} \right)^3
\end{eqnarray}
In all the above equation we introduce an arbitrary mass scale $m_{\phi}$ for convenience in numerical computation. We will take it as the Hubble constant at the end of inflation. At this point, let us emphasize that our analysis is purely perturbative. However the non-perturbative effects may have an important role, we left it for our future studies. In order to solve, the above set of Boltzmann equation, we set the initial condition at the end of inflation as,
\begin{eqnarray}
\Phi(1) = \frac{3}{8\pi} \frac{M_{pl}^2 H_I^2}{m_{\phi}^4};~~~~~R(1)=X(1)=0. 
\label{PhiI}
\end{eqnarray}
Where the initial Hubble expansion $H_I^2 = (8\pi/3M_{pl}^2)\rho_{\phi, end}$.
The set of Boltzmann equations can now be easily solved provided we know the inflaton decay with and the annihilation cross section.
The reheating temperature of the universe is identified at the instant of $H(t)=\Gamma_{\phi}$, when all the inflaton field energy is instantaneously converted into the radiation field. However, in general, this is not true. 
At any instant of time during the evolution, the temperature of our universe is identified with the radiation temperature with $T \equiv T_{rad}=\left[ {30}/({\pi^2 g_*)}\right]^{1/4} \rho_R^{1/4} $.
 The thermally averaged annihilation cross section of the dark matter from the radiation is $\left< \sigma v \right> $ \cite{Giudice:2000ex}. We will be using this as a free parameter in our subsequent numerical analysis.

\subsubsection{The early time solution and the maximum temperature for general $\omega_{\phi} = (n-2)/(n+2)$}
Before we resort to the numerical solution to find the connection among the CMB anisotropy and the dark matter abundance, let us compute an analytic expression for the maximum possible temperature during reheating phase for general equation of state of the inflaton field. In the early phase of the reheating stage ($H\gg\Gamma_{\phi} )$, considering the initial condition $R(A_I) \simeq X(A_I) \simeq 0$ with $A_I=1$, Eq.(\ref{Radeq}) can be solved as 
 \begin{eqnarray}
 \frac{dR}{dA} \simeq c_1 (1+w_{\phi}) \Phi_I^{\frac{1}{2}} A^{\frac{3}{2}(1-w_{\phi})}
 \implies R    \simeq 2 c_1\left(\frac{1+\wI}{5-3\wI} \right) \left[A^{\frac{5-3\wI}{2}}  - A_I^{\frac{5-3\wI}{2}}\right].
\end{eqnarray}
Therefore, by using the above solution and using the definition of temperature in terms of the radiation density, one arrive at the following expression
\begin{eqnarray}
 T = \left[ \frac{60 c_1}{g_* \pi^2}~ \frac{1+\wI}{5-3\wI}\right]^{\frac{1}{4}} ~m_{\phi} \left( \frac{\Phi_I}{A_I^{3(1+\wI)}}\right)^{\frac{1}{8}} \left[ \left(\frac{A}{A_I}\right)^{-\frac{3}{2}(1+\wI)}  - \left(\frac{A}{A_I}\right)^{-4}\right]^{\frac{1}{4}}
 \label{temp1}
\end{eqnarray}
By using the the eq.(\ref{PhiI}), the maximum temperature can be found as 
\begin{align}
\nno
T_{max} = \left[ \frac{60 c_1}{g_* \pi^2}~ \frac{1+\wI}{5-3\wI}\right]^{\frac{1}{4}} ~&m_{\phi}^{\frac{1}{2}} \left(\frac{3}{8\pi} \frac{M_{pl}^2 H_I^2}{A_I^{3(1+\wI)}}\right)^{\frac{1}{8}}\times \\ \phantom{\times}&\left[ \left( \frac{8}{3(1+\wI)}\right)^{-\frac{3(1+\wI)}{5-3\wI}} - \left( \frac{8}{3(1+\wI)}\right)^{-\frac{8}{5-3\wI}} \right]
\end{align}
In the following numerical analysis for the dark matter abundance, we will find maximum temperature plays an important role in constraining the dark matter parameter space. The essential idea behind this temperature is that as the reheating temperature is measured at a later stage of reheating, the radiation production commences at the very early stage. As one can clearly see the non-trivial dependence of the temperature on the inflaton equation of state parameter $(\omega_{\phi})$ and the inflaton decay constant $\Gamma_{\phi}$.  
 Depending upon the initial value of Hubble rate which has a direct connection with the CMB anisotropy, the maximum temperature can be many order higher than the reheating temperature\cite{Kolb:1990vq,Chung:1998rq,Giudice:2000ex}. Therefore, this maximum temperature will play as an intermediate scale between the inflationary energy scale and the reheating temperature. Subsequently we show how this will effect the dark matter production mechanism during depending upon the dark matter mass.
 
\subsubsection{Dark matter relic abundance and its $(T_{re}, T_{max})$ dependence}
As we have emphasized already, we will describe how the present dark matter abundance is controlled by the CMB anisotropy through the inflationary model and its subsequent reheating phase. The present dark matter abundance parametrized by the normalized density parameter $\Omega_X$ can be expressed in terms of present day radiation abundance $\Omega_R$ $(\Omega_R h^2=4.3\times10^{-5})$, as follows
\begin{eqnarray}
\Omega_X h^2 = \left<E_X\right> \frac{X(T_F)}{R(T_F)} \frac{T_F}{T_{\rm now}} \frac{A_F}{m_{\phi}} \Omega_R h^2	
\label{eq:relicX}
\end{eqnarray}
Where, $T_F$ is the temperature at very late time when dark matter and radiation comoving energy densities became constant. The current CMB temperature is given by $T_{\rm now} = 2.35\times10^{-13}{\rm GeV}$. 
For our subsequent discussions, it is important to know the behavior of the dark matter abundance in terms of reheating parameters $(T_{\rm re},M_X, \langle \sigma v\rangle)$. The analytic expressions for the dark matter abundance for different dark matter mass $M_X$ can be calculated following the references \cite{Giudice:2000ex}. As we have considered a generic equation of state parameter for the inflaton, we are interested in the generalized expression for dark matter abundance for arbitrary $\wI$. {The dependence of relic abundance on reheating temperature, the dark matter mass, and annihilation cross section can be found out by tracking the early time behavior of the evolution equation for X in Eq.\eqref{Xeq}. In the early times we take $\Phi = \Phi_{\rm I}$, $R=0$, and consider
$X \ll X_{EQ}$, thus, the early time behavior of $X$ can now be written as
\be
X' ~=~ c_2 \Phi^{-\frac{1}{2}}_{\rm I} A^{-\frac{(5-3w)}{2}} X^2_{\rm EQ}
\label{x-early}
\ee
the equilibrium value $X_{eq}$ is given in terms of temperature which can be found from early time solution of $R$.
The solution of integral in Eq.(\ref{x-early}) can be found by approximating it as a Gaussian integral\cite{Chung:1998rq,Giudice:2000ex}. Following the approach given in \cite{Giudice:2000ex}, for general $\wI$, we can find out the final X-density $X(\infty)$ which when substituted back in \eqref{eq:relicX} fetch us with the following relations:} 
\bea
\Omega_X h^2 &\propto&  \left<\sigma|v|\right> M_X^4 \exp\left[-\frac{ (17+w)}{ 2(1+w)} \frac{M_X}{T_{\rm max}}\right]~~ \text{for $M_X \gtrsim T_{\rm max}$} \label{eq28}\\
	\Omega_X h^2 &\propto& \left< \sigma v\right> \frac{T_{re}^{\frac{7+3\wI}{1+\wI}}}{M_X^{\frac{9-7\wI}{2(1+\wI)}}} \qquad \text{for~} T_{\rm max} > M_X > T_{\rm re}.
	\label{eq29}
\eea
One can particularly notice the non-trivial dependence of $w_{\phi}$ for two different dark matter mass range with respect to the $T_{max}$ as derived before. The proportionality factors depends on other inflationary parameters. As we are solving the equations numerically to find the final $X$ density, the exact form of these factors will not be important for our subsequent discussions. Furthermore, if we consider the dark matter mass to be less than the reheating temperature, the \textit{freeze-in} happens after the reheating is over. For that case the relic abundance can be expressed as \cite{Dev:2013yza}
\bea
\Omega_X h^2 \propto \left< \sigma v\right> M_X~T_{\rm re} \qquad \text{When~}M_X < T_{\rm re},
\label{eq30}
\eea
For all the above three expressions for the dark matter abundance, we have considered the freeze-in mechanism. This essentially means that dark matter will never reach the equilibrium with the thermal bath. In the literature, these dark matter is known as feebly interacting dark matter(FIMP)\cite{fimp}. {We also note that the freeze-in dark matter production during reheating is sensitive to the early history before the decoupling era(UV dominated) which essentially the case of DM production through heavy mediators\cite{Mambrini:2013iaa,Chu:2013jja,Blennow:2013jba,Elahi:2014fsa,Mambrini:2015vna,Nagata:2015dma,Chen:2017kvz,Bernal:2018qlk,Garcia:2018wtq}.} Following our recent work\cite{Maity:2018dgy}, in the next section, we will briefly outline the steps to connect CMB, reheating and dark matter abundance.

\subsection{Connection between CMB and dark matter: Methodology}
In this section we will briefly review the methodology developed in \cite{Maity:2018dgy} which describes the connection between the CMB anisotropy and the dark matter phenomenology.
For the canonical inflation, the important physical quantities are  $(N_k, H_k, V_{\rm end}(\phi_k))$ which we have already defined. Since we would like to compute for a specific CMB  scale $k$ (CMB pivot scale), all those quantities are expressed as  
\begin{equation}
H_k = \frac{\pi M_p \sqrt{r A_s}}{\sqrt{2}}~~
;~~	N_k = ln\left(\frac{a_{\rm end}}{a_k} \right) =  \int_{\phi_k}^{\phi_{\rm end}} \frac{1}{\sqrt{2\epsilon_V}} \frac{|d\phi|}{M_p} .
\label{Hk}
\end{equation}
where $(\phi_{\rm end}$ is computed form the condition of end of inflation
\be
\epsilon(\phi_{\rm end}) = \frac{M_{\rm p}^2}{2} \left(\frac{V'(\phi_{\rm end})}{V(\phi_{\rm end})}\right)^2 =1,
\ee
While, the value of the scalar spectral index $n_s^k$ at the horizon crossing of a particular scale $k$ is
\bea
n^k_s = 1 - 6\epsilon(\phi_k) + 2\eta(\phi_k)
\eea
Now, the above expression can be inverted to find $\phi_k$ in terms of the scalar spectral index. This relation provides us the connection between CMB anisotropy and the inflationary dynamics. Thus using this constraint from CMB on inflation, we will set the initial condition for the subsequent reheating dynamics. Next, we will solve the Boltzmann Eqs.(\ref{Inflatoneq}-\ref{Xeq}) for all the three components of energy densities and also the scale factor $a$ during the reheating phase. For numerical analysis one of the important quantities called reheating e-folding number $N_{\rm re}$ measuring the duration of reheating is computed from the standard definition $N_{\rm re} = \ln(a_{\rm re}/a_{\rm end})$. Where $a_{\rm re}$ is the scale factor at the end of reheating which is defined as 
\bea
H(a_{re})^2= \frac{\dot{a}_{\rm re}}{a_{\rm re}} =\frac{8\pi}{3M_{\rm Pl}^2} (\rho_\phi(\Gamma,n_s^k,M_X) +\rho_R(\Gamma,n_s^k,M_X) +\rho_X(\Gamma,n_s^k,M_X)) = \Gamma^2 .
\label{recond}
\eea
Where each of the energy components is written as an implicit function of reheating parameters and inflationary power spectrum $n_s^k$ as they are the solution of the Boltzmann equations with the initial condition at the end of inflation expressed in eq.(\ref{PhiI}). 
This is also the scale factor at which we define the radiation temperature as another important parameter called reheating temperature with the following relation
\bea 
T_{\rm re} \equiv T^{\rm end}_{\rm rad}=\left[ {30}/{\pi^2 g_*(T)}\right]^{1/4} \rho_R(\Gamma,n_s,M_X)^{1/4}.
\eea  
At this point, it is important to note that in the present case the reheating temperature cannot be straightforwardly expressed in terms of inflaton decay width $\Gamma$, as can be clearly seen from the above equation. Never the less, the cosmological scale $k$ observed in CMB is originated during inflation at the horizon scale $H_k$, and then evolved through the subsequent reheating and radiation phase. Therefore, an important relation has been can established among  $(H_k, T_{re}, N_{re}, T_0)$ in  \cite{Dai:2014jja}, as
\begin{equation}
T_{\rm re} = \left(\frac{43}{11 g_{\rm re}}\right)^{\frac{1}{3}}  \left(\frac{a_0 T_0}{k} \right) H_k e^{-N_k} e^{-N_{\rm re}} .
\label{TreEq}
\end{equation}
where $T_0 = 2.725{\rm K}$ is the present CMB temperature, $g_{re}$ is the effective number of light species, and $H_0$ is the present value of the Hubble parameter. Hence one gets an important connection among the CMB anisotropy at a particular scale $k$, inflation and the reheating period through the following equation,
\bea
\left(\frac{43}{11 g_{\rm re}}\right)^{\frac{1}{3}}  \left(\frac{a_0 T_0}{k} \right) H_k e^{-N_k} e^{-N_{\rm re}} =\left[ {30}/{\pi^2 g_*(T)}\right]^{1/4} \rho_R(\Gamma,n_s,M_X)^{1/4} 
\eea
However, following the reference \cite{Maity:2018dgy}, in addition to radiation, we also considered the production of dark matter. Therefore, 
the observation of present dark matter abundance $\Omega_X h^2 = 0.12$ will lead to the following another important connection between CMB anisotropy, reheating and the dark matter phenomenology. Following the relations given in Eqs.(\ref{eq28},\ref{eq29},\ref{eq30}), the aforementioned connection are expressed as follows,
\bea \label{abundance}
&&\Omega_X h^2 \propto  \left<\sigma|v|\right> M_X^4 \exp\left[-\frac{(17+w)M_X}{(1+w)T_{\rm max}}\right]~~ \text{for $M_X \gtrsim T_{\rm max}$} \nno\\
&&\Omega_X h^2 \propto \left< \sigma v\right> \frac{T_{re}^{\frac{7+3\wI}{1+\wI}}}{M_X^{\frac{9-7\wI}{2(1+\wI)}}} \propto  \frac{\left< \sigma v\right>}{M_X^{\frac{9-7\wI}{2(1+\wI)}}} \left[\left(\frac{a_0 T_0}{k} \right) H_k e^{-N_k} e^{-N_{\rm re}}\right]^{\frac{7+3\wI}{1+\wI}}\qquad \text{for~} T_{\rm max} > M_X > T_{\rm re} \nno\\
&&\Omega_X h^2 \propto \left< \sigma v\right> M_X~T_{re} \propto  \left< \sigma v\right> M_X \left(\frac{a_0 T_0}{k} \right) H_k e^{-N_k} e^{-N_{re}} \qquad \text{When~}M_X < T_{\rm re}.
\eea 
Depending upon the mass of the dark matter we can clearly see the behavior of the dark matter parameter space intimately connected to the CMB scale and the associated temperature.  
To this end let us also count the number of independent parameter of our study. At the current epoch of our universe, we have two observables corresponding to the CMB anisotropic power spectrum and the dark matter abundance. Based on our assumption that the radiation and the dark matter are produced through inflaton decay, we have four parameters $(\phi_*, \langle \sigma v\rangle,\Gamma, M_X)$ during reheating. Therefore, given the initial condition set by the CMB on the inflaton and present dark matter abundance, we have two independent parameters $(\phi_*, M_X)$ left.  
With this strategy, we will study the Boltzmann equations during reheating and show how for a given set $(\phi_*,M_X)$, we can constrain the dark matter annihilation cross-section through the CMB anisotropy. 

\subsubsection{Numerical results, constraints on the model}

With the help of formalism described above, in this section, we would like to describe the constraints on various parameters of our model. As we have already discussed, for a particular model with $n$ value, we have two independent parameters $(\phi_*,M_X)$. The first and foremost point we would like to make is that the production of dark matter does not have much effect on the determination of reheating temperature for all the models. However it is important to mention that specifically for the dark matter mass $M_X > T_{\rm re}$, it is the reheating period which plays important role in determining the present dark matter abundance. 
In the following we have considered $n=2,4,6,8$ and three sample values of $\phi_* = 10,1,0.1$ in unit of Planck mass. Before we go to
the quantitative discussion, let us point out the general results for all the models and there generic behavior for different parameters values. 
For the class of minimal models we discussed, the spectral index is expressed as
\begin{equation}
n^k_s \simeq 1 - \frac{2n(n+1)M_p^2}{\phi_*^2} \frac{1}{\tilde{\phi}_k^{n+2}}
\end{equation}
Where, $\phi_k$ is the field value of a mode $k$ at horizon crossing. This equation can be inverted to find $\phi_k$ in terms of $n_s$. Once, we know this relation all the quantities such as tensor-to-scalar ratio $r$, the efolding number between the horizon crossing of a particular mode $k$ and the end of inflation $N_k$ and the parameter $m$(for $n \neq 4$) or $\lambda$(for $n = 4$) can be found in terms of $n_s$ and $\phi_*$. 

\begin{figure}[t]
	\centering	
	\subfigure[]{\includegraphics[scale=0.4]{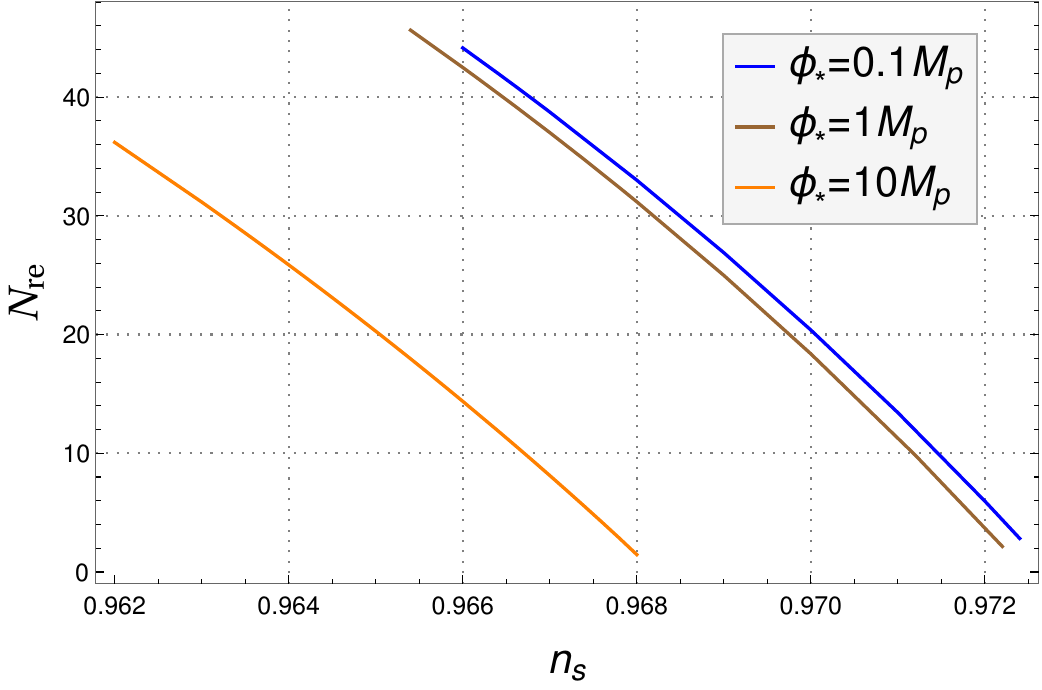}}
	\subfigure[]{\includegraphics[scale=0.4]{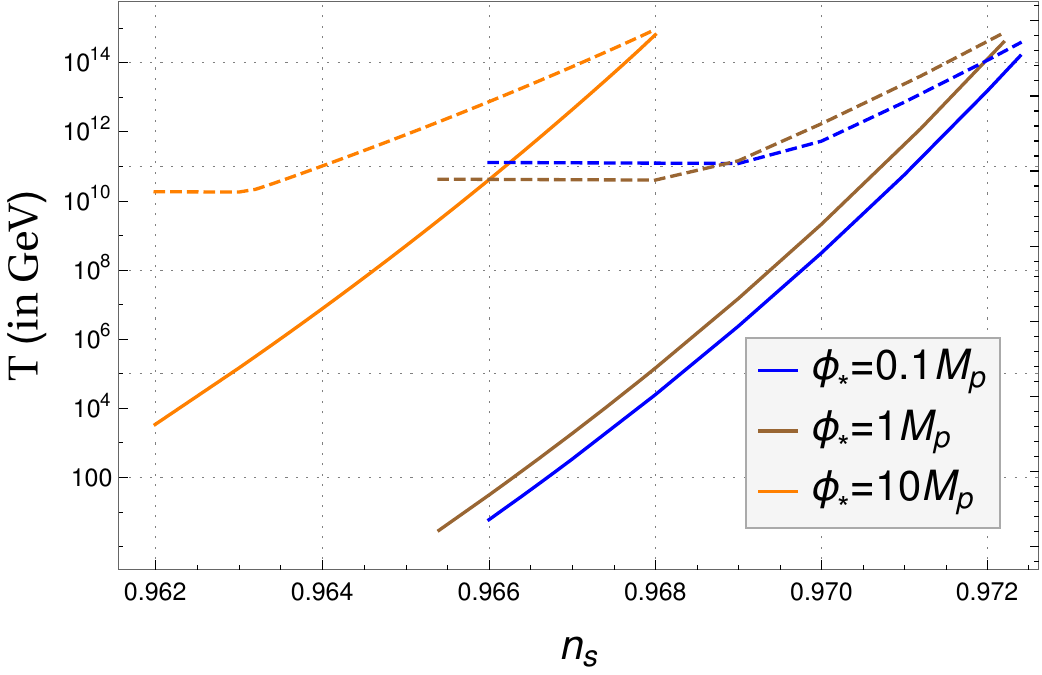}}
	\caption{Evolution of (a) Reheating e-folding number $N_{re}$ during reheating and (b) the reheating temperature $T_{re}$(solid lines) and the maximum temperature $T_{max}$(dotted lines) with respect to $n_s$ for $n=2$ for three different values of $\phi_*$. If one extrapolates the solid and dotted lines, one will have maximum reheating temperature $T_{re}^{max}\simeq 10^{15}$ GeV. All the plots are independent of dark matter masses for a set of given initial condition.}
	\label{n2_nsnret}
\end{figure}

For $n<4$, in $(n_s~vs~N_{\rm re})$ plot, our generic observation is that with the increasing $n_s$, the reheating efolding number $N_{\rm re}$ is decreasing as seen in fig.(\ref{n2_nsnret}a), and as a consequence reheating temperature $T_{re}$ is increasing shown in fig.(\ref{n2_nsnret}b). This behavior suggest  the existence of maximum reheating temperature $T_{\rm re}^{\rm max}$ thereby providing the maximum allowed scalar spectral index $n_s^{\rm max}$. This also indicates the instantaneous reheating for $n_s = n_s^{\rm max}$. On the other hand for $n>4$, what we observed is the opposite that is the reheating efolding number $N_{\rm re}$ is increasing as seen in fig.(\ref{n6_nsnret}a), and as a consequence reheating temperature $T_{\rm re}$ is decreasing with $n_s$, as shown in fig.(\ref{n6_nsnret}b). The maximum reheating temperature corresponding to a minimum possible $n_s^{min}$. In this case the instantaneous reheating occurs for $n_s = n_s^{min}$. This feature also have been observed in work \cite{Dai:2014jja}, with the effective single fluid equation of state formalism. 
However, another important fact that we observed is the existence of possible maximum reheating temperature $T_{\rm re}^{\rm max}\simeq 10^{15}$ GeV irrespective the of different parameters in our models. For all values of $n$ and $\phi_*$, we found this temperature has an important connection with the CMB anisotropy. As we have already described in detail, by tuning the value of $\phi_*$, we can go from large field to small field inflation model. However, our prediction of $T_{\rm re}^{\rm max}$ appeared to be independent of the inflation model. We will elaborate on this issue in our future publication.        
    Never the less, from the numerical fitting we express the reheating temperature $T_{\rm re}$ in terms of spectral index $n_s$ with the following approximate relation, 
\bea
\log_{10}\left( T_{\rm re} {~\rm GeV}\right) \simeq Q_p\left[A + B(n_s - 0.962) + C(n_s - 0.962)^2 \right]. 
\label{nsTrefor}
\eea
Where, the dimensionless constants $A=4$, $B =1.5\times10^3$ and $C=6\times 10^4$ for $n=2$, $A=20$, $B =-5\times10^3$ and $C=-1.5\times 10^4$ for $n=6$ and $A=20$, $B =-3\times10^3$ and $C=-1.5\times 10^4$ $n=8$. The proportionality constant $Q_p$ is $\phi_*$ dependent constant.
 From the above eqs.(\ref{nsTrefor}) and (\ref{abundance}), we can clearly build an important connection between the dark matter abundance and the CMB anisotropy which is directly connected to the scalar spectral index $n_s$. 
In the next section, we will present our detail numerical results.
\subsubsection{Results and constraints: {\bf Model n=2}}
This is similar to the usual quadratic chaotic model. However, as we have seen before the new parameter $\phi_*$ controls the behavior of cosmological parameters in a significant way. From the Fig.(\ref{n2_nsnret}.a-b) we see the behavior  of $N_{re}$ and $(T_{re}, T_{max}$) in terms of $n_s$. It is evident that the allowed region of $n_s$ shifts towards lower values as we increase $\phi_*$, which in fact resembles the usual chaotic inflation. For $\phi_* < M_p$ we have small field inflation and the values of $(n_s,r)$ are in agreement with the observation. 
\begin{figure}[t]
	\centering
	\subfigure[]{\includegraphics[scale=0.3]{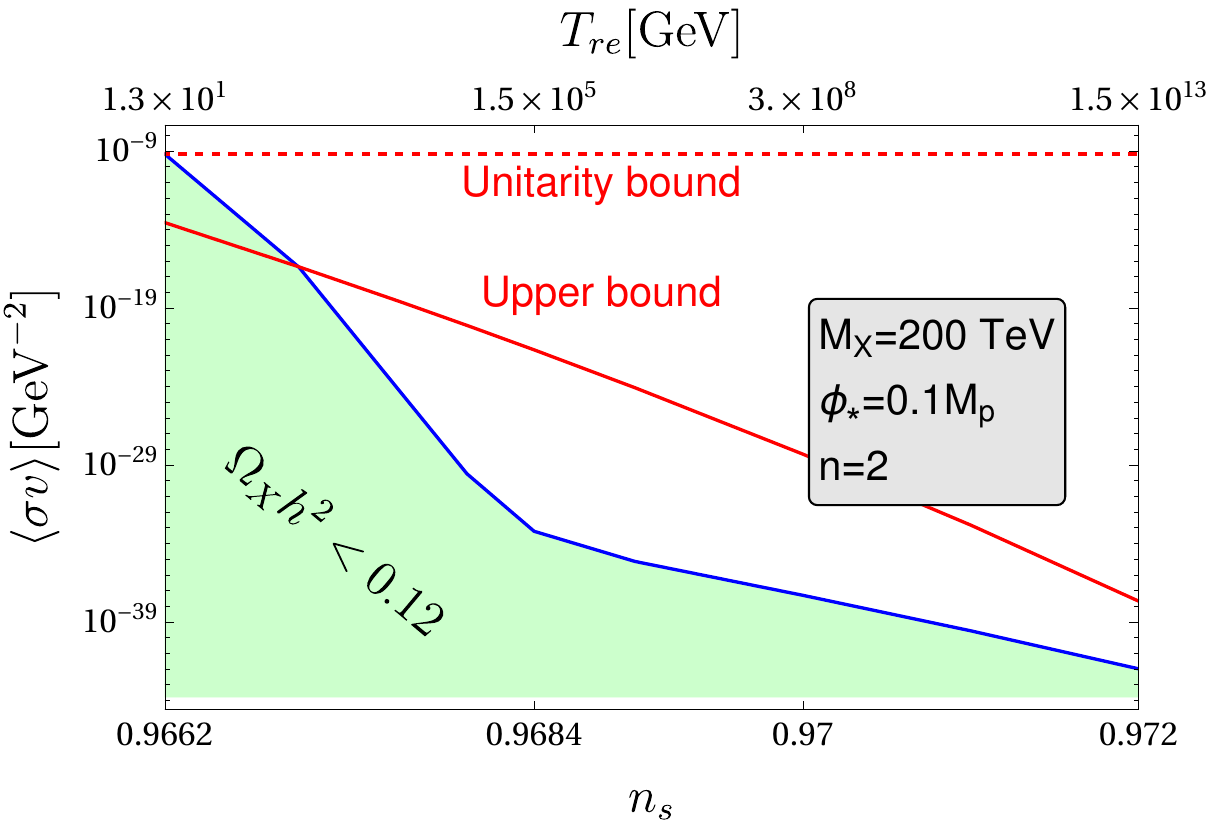}}
	\subfigure[]{\includegraphics[scale=0.3]{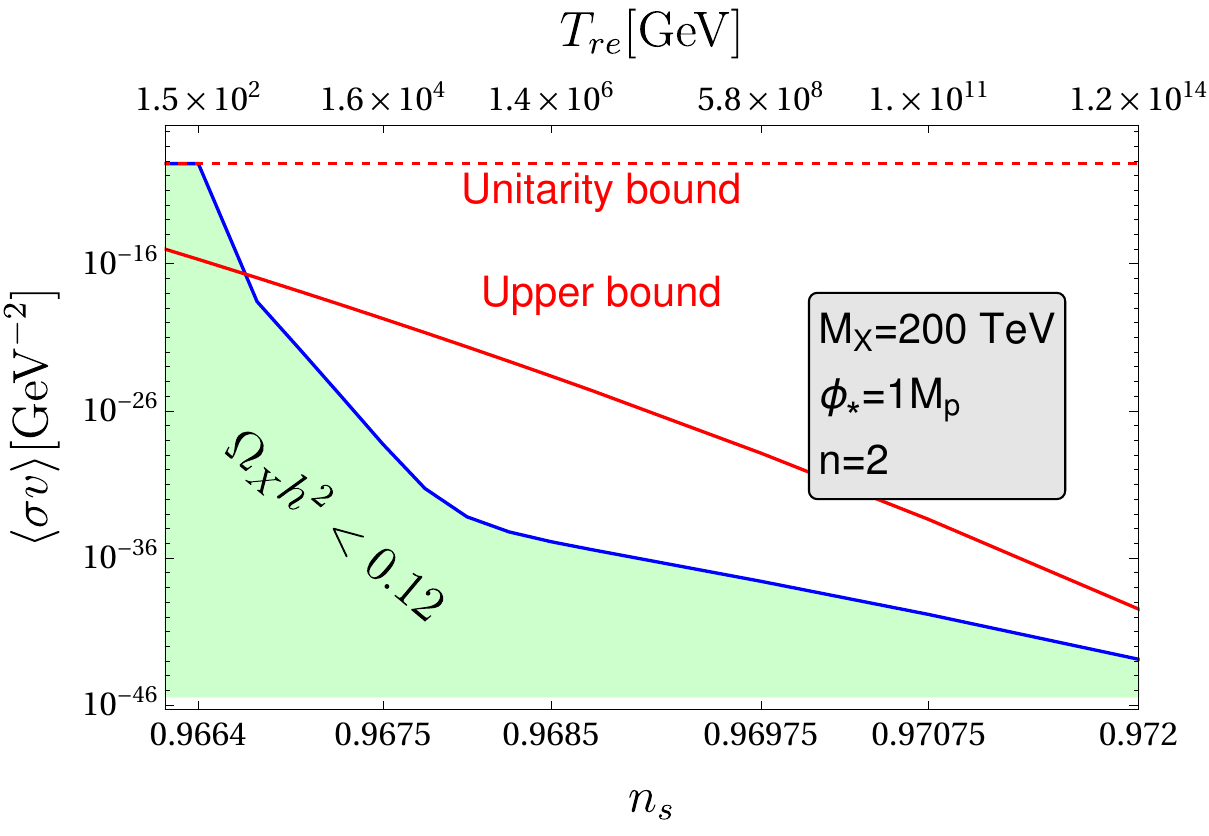}}
	\subfigure[]{\includegraphics[scale=0.3]{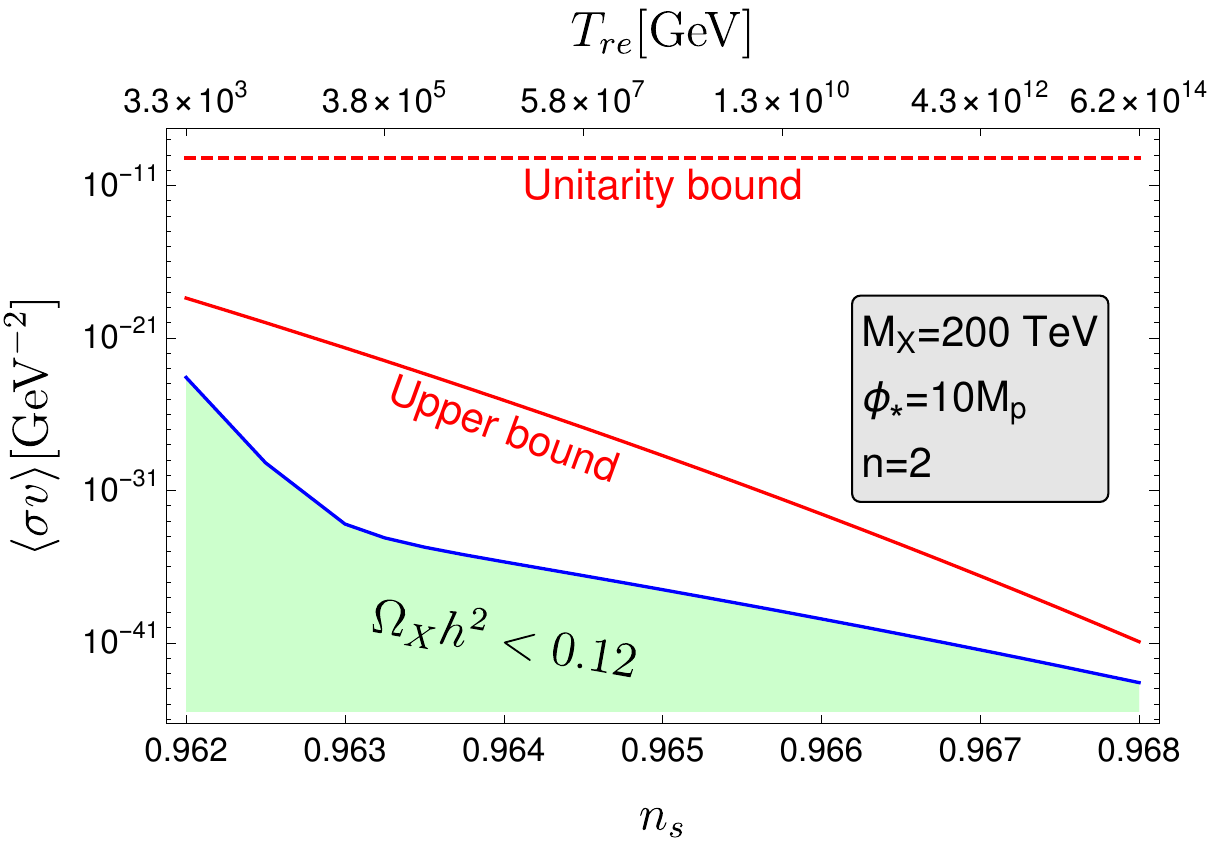}}
	\caption{$(\langle\sigma v\rangle~vs~n_s)$ were plotted for three values of $\phi_*$ and $n=2$. The blue line with green shaded region corresponds to dark matter abundance $\Omega_X h^2 \leq 0.12$. We consider three possible values of $\phi_*$. For all the cases we have chosen a dark matter mass $M_X=200 TeV$. Red horizontal line corresponds unitarity bound described in the main text.}
	\label{n2nssig}
\end{figure}
\begin{figure}[t]
	\centering
	\subfigure[]{\includegraphics[scale=0.3]{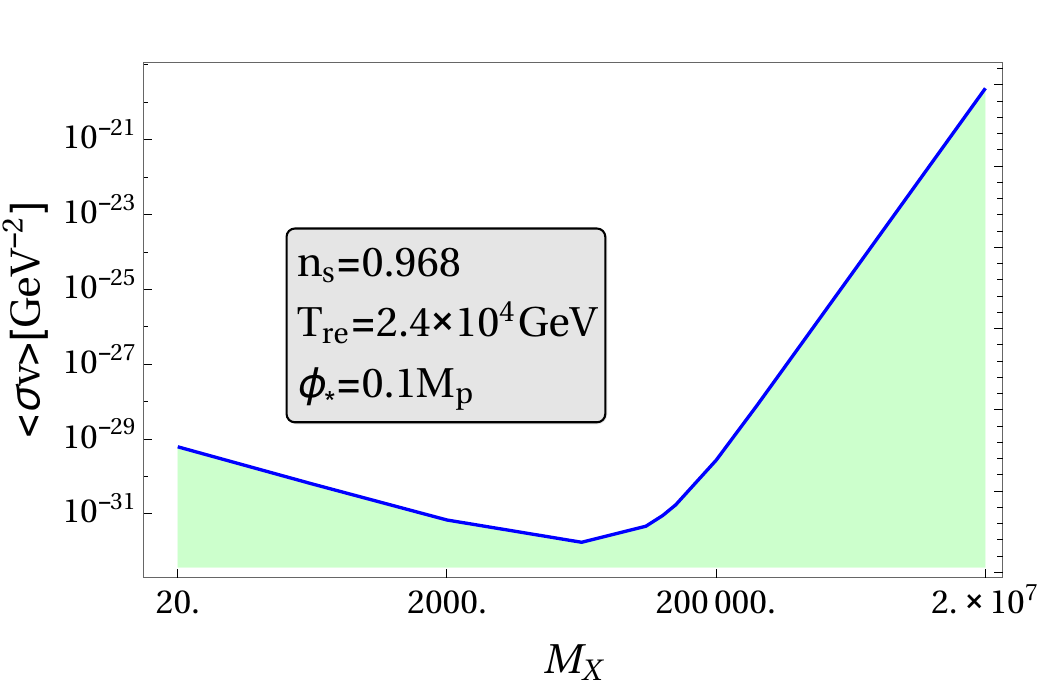}}
	\subfigure[]{\includegraphics[scale=0.3]{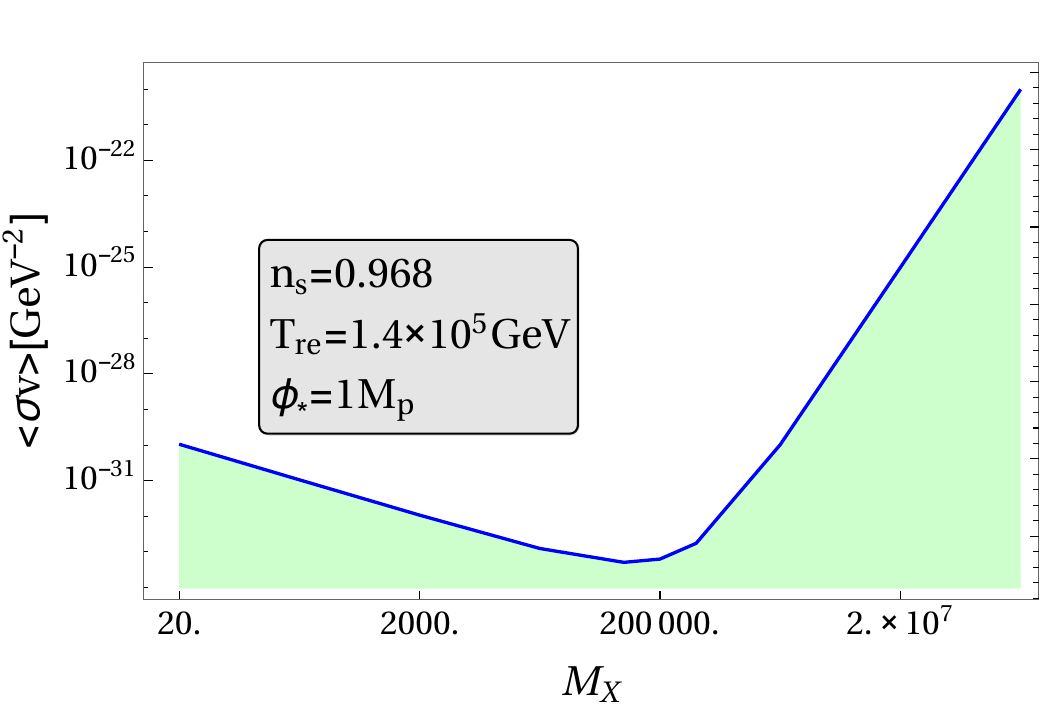}}
	\subfigure[]{\includegraphics[scale=0.3]{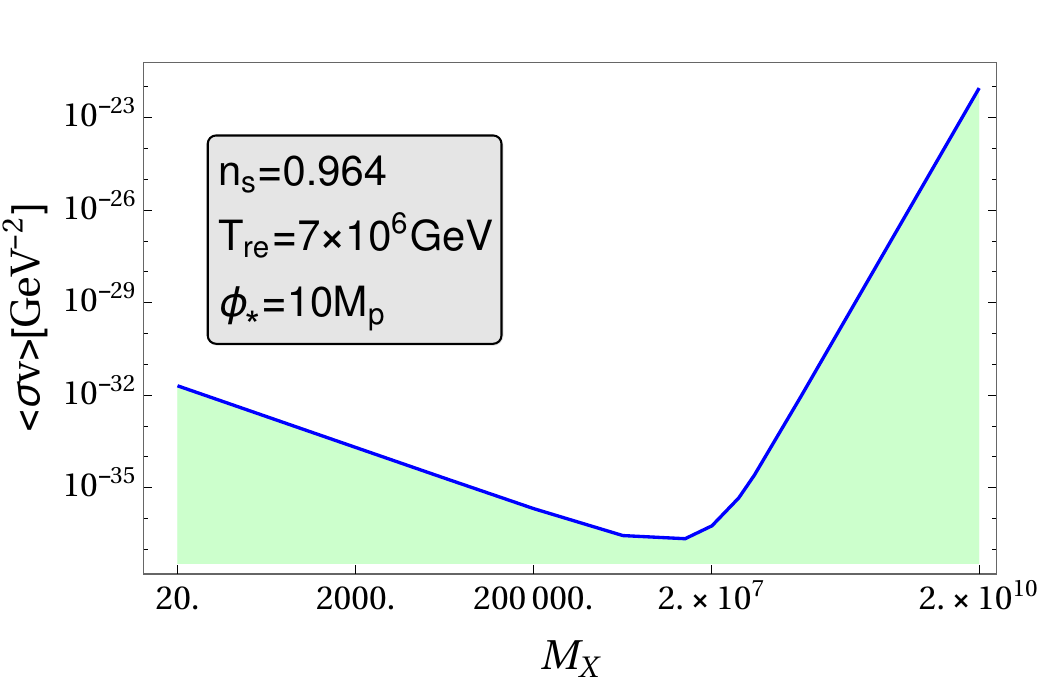}}
	\caption{$(\langle\sigma v\rangle~vs~M_X)$ were plotted for three values of $\phi_*$ and $n=2$. Depending upon the available ranges of temperature within the $1\sigma$ range of $n_S$, we have considered three possible values of $n_s$. Minimum of the blue curve corresponds to the equality $M_X = T_{re}$. It is obvious that at this point the required value of the cross-section is minimum for a given abundance $\Omega_X h^2 = 0.12$}
	\label{n2mxsig}
\end{figure}
 \begin{table}[h!]
	\label{modeln2}
	Model parameters and associated constraints on dark matter parameters for n=2
	\begin{minipage}{.33\linewidth}
		\caption{$\phi_*=0.1M_p$}
		\begin{tabular}{|c|c|c|c}\hline
			{$n_s$} & {$N_{re}$} & {$T_{re}(GeV)$} & $\begin{array}{c}{\fontsize{7}{7}\selectfont \textit{$\langle \sigma v\rangle GeV^{-2}$}}\\ {\fontsize{7}{7}\selectfont \textit{$M_X(200 TeV)$}}\end{array}$\\\hline
			{0.966} & {44} & {$5.8$} & {\textit{unitarity}}\\
			{0.968} & {33} & {$2.4\times 10^{4}$}   & {$3\times 10^{-30}$}\\
			{0.970} & {20} & {$3\times 10^8$}       & {$5\times 10^{-38}$}\\
			{0.972} & {6}  & {$1.5\times 10^{13}$} &  {$1\times 10^{-42}$}\\\hline
		\end{tabular}
	\end{minipage}%
	\begin{minipage}{.33\linewidth}
		\caption{$\phi_*=1M_p$}
		\begin{tabular}{|c|c|c|c}\hline
			{$n_s$} & {$N_{re}$} & {$T_{re}(GeV)$} & $\begin{array}{c}{\fontsize{7}{7}\selectfont \textit{$\langle \sigma v\rangle GeV^{-2}$}}\\ {\fontsize{7}{7}\selectfont \textit{$M_X(200 TeV)$}}\end{array}$\\\hline
			{0.966}  & {42.5} & {$30.1$}             & {\textit{unitarity}}\\
			{0.968}  & {31}   & {$1.4\times 10^5$}   & {$6\times 10^{-34}$}\\
			{0.970}  & {18}   & {$2\times10^9$}      & {$2\times10^{-40}$}\\
			{0.9722} & {2}    & {$3.8\times10^{14}$} & {$3\times10^{-44}$}\\\hline
		\end{tabular}
	\end{minipage}%
	\begin{minipage}{.32\linewidth}
		\caption{$\phi_*=10M_p$}
		\begin{tabular}{|c|c|c|c|}\hline
			\textit{$n_s$} & \textit{$N_{re}$} & \textit{$T_{re}GeV$} & $\begin{array}{c}{\fontsize{7}{7}\selectfont \textit{$\langle \sigma v\rangle GeV^{-2}$}}\\ {\fontsize{7}{7}\selectfont \textit{$M_X(200 TeV)$}}\end{array}$\\\hline
			{0.962} & {36} & ${3.3\times 10^{3}}$ & {$3\times10^{-24}$}\\
			{0.964} & {26} & ${7.4\times 10^6}$   & {$2\times 10^{-36}$}\\
			{0.966} & {14} & ${4\times 10^{10}}$  & {$4\times10^{-40}$}\\
			{0.968} & {1.5} & ${6\times10^{14}}$  & {$3\times10^{-44}$}\\\hline
		\end{tabular}
	\end{minipage}%
\end{table}
For these small field models, therefore, within the $1\sigma$ range of $n_s$,  very small reheating temperature can be achieved. For example at the central value of $n_s=0.968 $, reheating temperature turned out to be $10^4\sim 10^5$ GeV for $\phi_*=(1,0.1)M_p$, whereas for $\phi_*=10 M_p$, reheating temperature is very high $\sim 10^{14}$ GeV. These fact can be seen from sample values given in the Tab.\ref{modeln2} and also from the Figs.(\ref{n2_nsnret}.a-b). The lower limit of the value of $n_s$ is set from BBN constraints\cite{Kawasaki:1999na,Kawasaki:2000en,Steigman:2007xt,Fields:2014uja}, which is $T_{re} \sim 0.1GeV$. On the other hand, we can see from the all the plots as well as from the data tables, there exists a maximum temperature(corresponding to the instantaneous reheating) irrespective of the value of the scale $\phi_*$. 


In addition to the one to one correspondence between $n_s$ and $T_{re}$, we simultaneously get constraints on the dark matter parameter space as shown in fig.(\ref{n2nssig}), and fig.(\ref{n2mxsig}). We have chosen a sample dark matter mass $M_X=200 TeV$. {This value of mass, throughout all the models, is chosen such that it corresponds to the reheating temperature for $n_s \sim 0.968$ for the intermediate value of the scale $\phiast$ among the three chosen values i.e., $\phiast = 1{\rm M_p}$.} It is clear that once $n_s$ is fixed, for a given $(\phi_*, M_X$), the scattering cross-section is also fixed. Therefore, within the $1\sigma$ range of $n_s$, the dark matter cross-section $\langle \sigma v\rangle$ is bounded given the value of dark matter abundance $\Omega_X h^2 = 0.12$. More importantly depending upon the values of $(M_X, n_s)$ the freeze-in will occur in three different regimes as we have already described in detail and also can be observed in the change of slopes of blue curves as a function of $n_s$ as shown in Figs.(\ref{n2nssig}-\ref{n2mxsig}). From our analysis, we see that for a fixed dark matter mass, annihilation cross-section increases with decreasing $n_s$. 
{However, as the cross section can not be arbitrarily large, the perturbative unitarity limit on $\left<\sigma v\right>_{MAX} = 8\pi/M_X^2$\cite{Griest:1989wd} restricts the maximum cross section. In our case, we will also have another bound on the cross-section which could be understood as follows. The X-particle produced during reheating via Freeze-in is such that the particles will never reach thermal equilibrium. Most of the particle production happens around the temperature $T_{\ast} \simeq m_X/4$\cite{Giudice:2000ex}. The production is exponentially suppressed around this narrow region centered around $T_{\ast}$ Now for $T<T_{\ast}$, the total number of $X$-particle is frozen and their density is diluted by the expansion of the universe. The condition that $n_{X}(T=0)<n_X^{eq}(T_{\ast})$ will then implies an upper bound on the X-particle annihilation cross section $\langle\sigma v\rangle_{\ast}\equiv\langle\sigma v\rangle_{T=T_{\ast}}$ given as\cite{Giudice:2000ex,Fornengo:2002db}
\begin{equation}
 \langle\sigma v\rangle_{\ast} \leq 7\times10^{-14}\Big(\frac{2}{g}\Big)\Big[\frac{g_{\ast}(T_{\ast})}{10}\Big]\Big[\frac{10}{g_{\ast}(T_{\rm re})}\Big]^{\frac{1}{2}}\Big(\frac{M_X}{10 {\rm GeV}}\Big)\Big(\frac{100 {\rm MeV}}{T_{\rm re}}\Big)^2 {\rm GeV^2}.
 \label{eq:upplim}
\end{equation}
In the subsequent plots of cross section of the X-particle, we will show both of these bounds. The perturbative unitarity bound will be depicted as \textit{Unitarity bound} while the bound form the above equation will be denoted as \textit{Upper bound}.} From fig.(\ref{n2nssig}.a-b) when, $\phi_* =0.1M_p,~1M_p$ the production of dark matter particle of mass $200TeV$ for lower $n_s$ regime is forbidden due to the unitarity bound(shown as red dashed line) as well the upper limit(shown as red solid line). The upper described in (\ref{eq:upplim}) limit put stringent bound on $n_s^{\min}$ in this two cases. The lowest possible value of $n_s$ turns out to be around $0.967$. The highest value of $n_s$ is turned out be $n_s^{max}\simeq 0.972$ as noted before, depend upon the maximum attainable temperature for the model. This important constraint on the value of $n_s$ coming from dark matter sector could be very important to understand and needs further study. 
Figs.(\ref{n2mxsig}.d-f) illustrates the conventional behavior of annihilation cross-section in terms of dark matter mass. However, an important point we would like to make again is that depending upon the value of inflationary power spectrum $n_s$ or CMB anisotropy, we can shed light on the possible production mechanism given the value of dark matter mass. 

\subsubsection{Results and constraints: {\bf Model n=4}}
This is a very special case out of all the models. One of the main reasons is that for $n=4$, the inflaton behaves like radiation during reheating with the equation of state parameter $\omega_{\phi}=1/3$. It is known that during reheating if we consider the effective equation of state to be radiation like, then distinguishing the reheating period and the radiation dominated period becomes very difficult as can be seen from the schematic diagram fig.(\ref{scales}). This very fact makes the reheating parameters $(N_{re},T_{re})$ indeterminate which has been observed in \cite{Dai:2014jja}. Using the Boltzmann equations, too the above nature of problem come back but in a different form, as for $n=4$, the differential equation describing the evolution of inflation and radiation became identical. The above scheme of constraining the decay width from CMB constrains is not feasible.
\subsubsection{Results and constraints: {\bf Model n=6}}
As we know all the usual power law inflationary models are large field and consequently predicts large $r$ value. However, because of new scale $\phi_*$, we were able to make our model perfectly compatible with the observation for even stiffer model such as $n=6$ and $n=8$. 
\begin{figure}[h!]
	\centering	
	\subfigure[]{\includegraphics[scale=0.4]{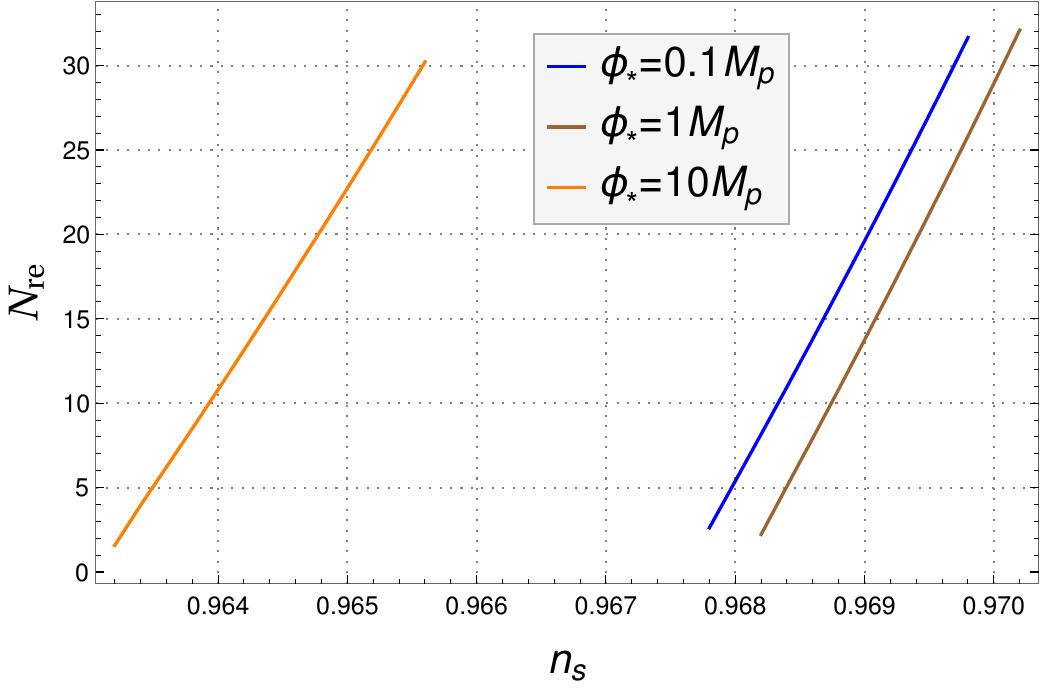}}
	\subfigure[]{\includegraphics[scale=0.4]{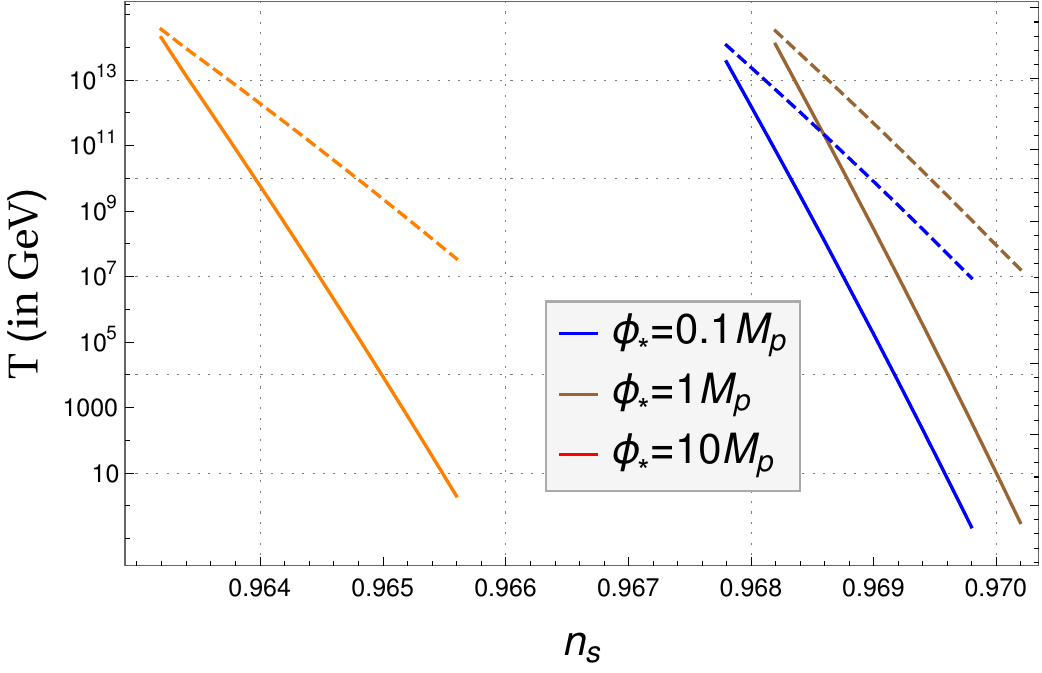}}
	\caption{Evolution of (a) Reheating e-folding number $N_{re}$ during reheating and (b) the reheating temperature $T_{re}$(solid lines) and the maximum temperature $T_{max}$(dotted lines) with respect to $n_s$ for $n=6$ for three different values of $\phi_*$. One clearly see the change of slop in compared to $n<4$ models. However, we observed maximum reheating temperature to the same $T_{re}^{max}\simeq 10^{15}$ GeV.}
	\label{n6_nsnret}
\end{figure}
\begin{figure}[t]
	\centering	
	\subfigure[]{\includegraphics[scale=0.3]{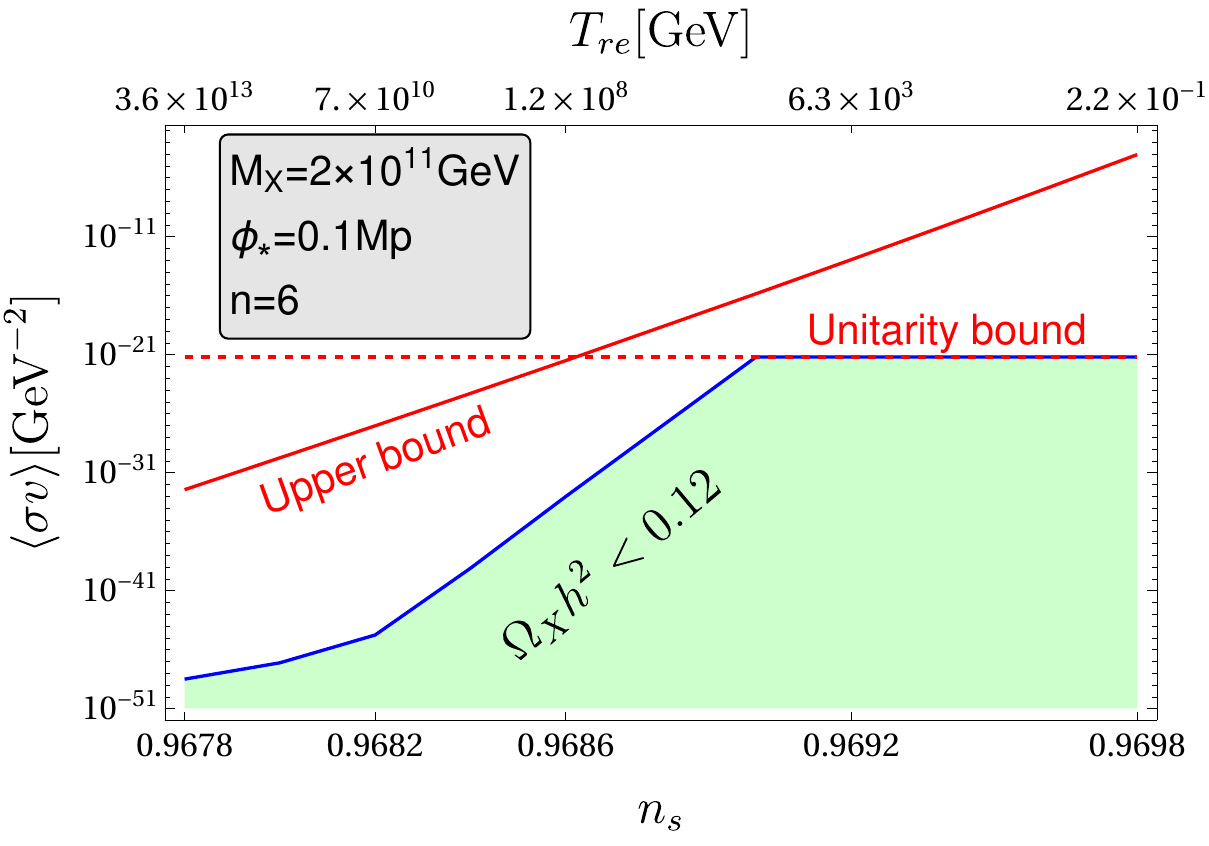}}
	\subfigure[]{\includegraphics[scale=0.3]{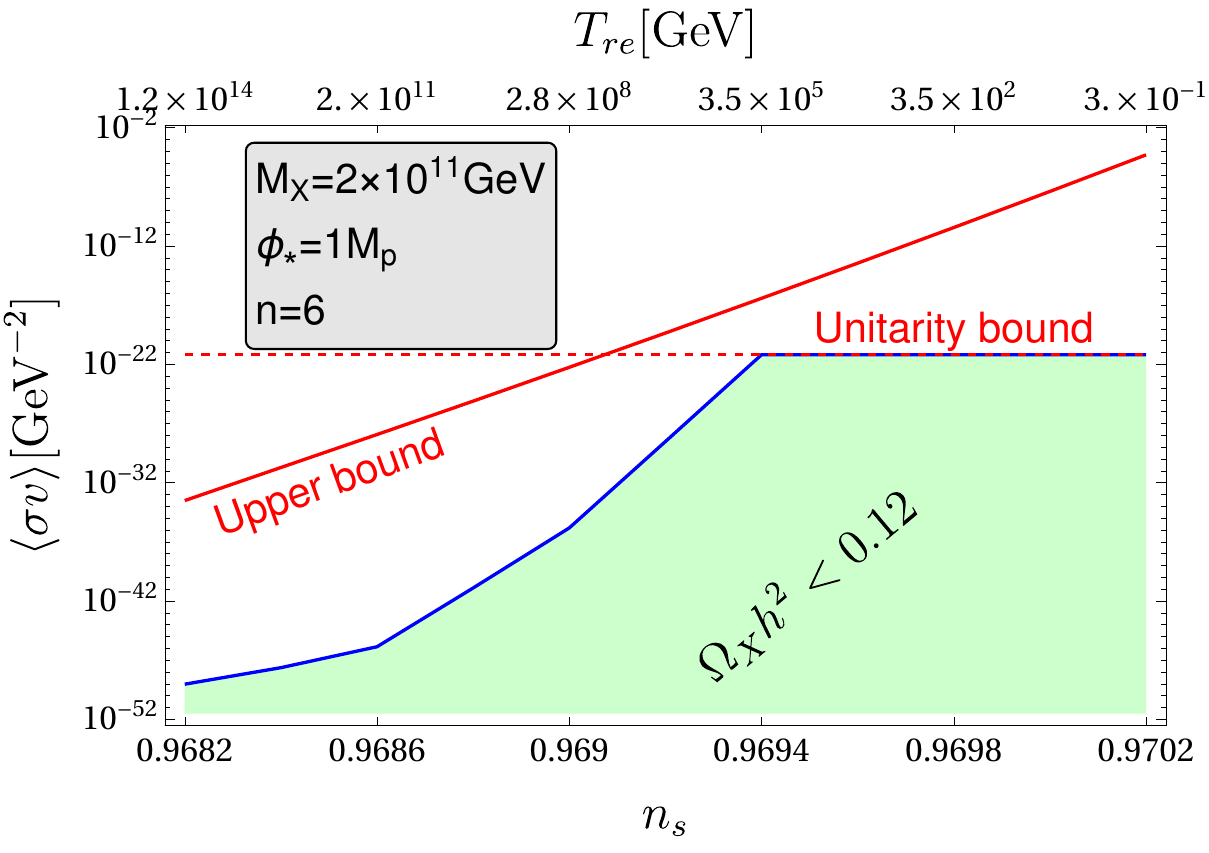}}
	\subfigure[]{\includegraphics[scale=0.3]{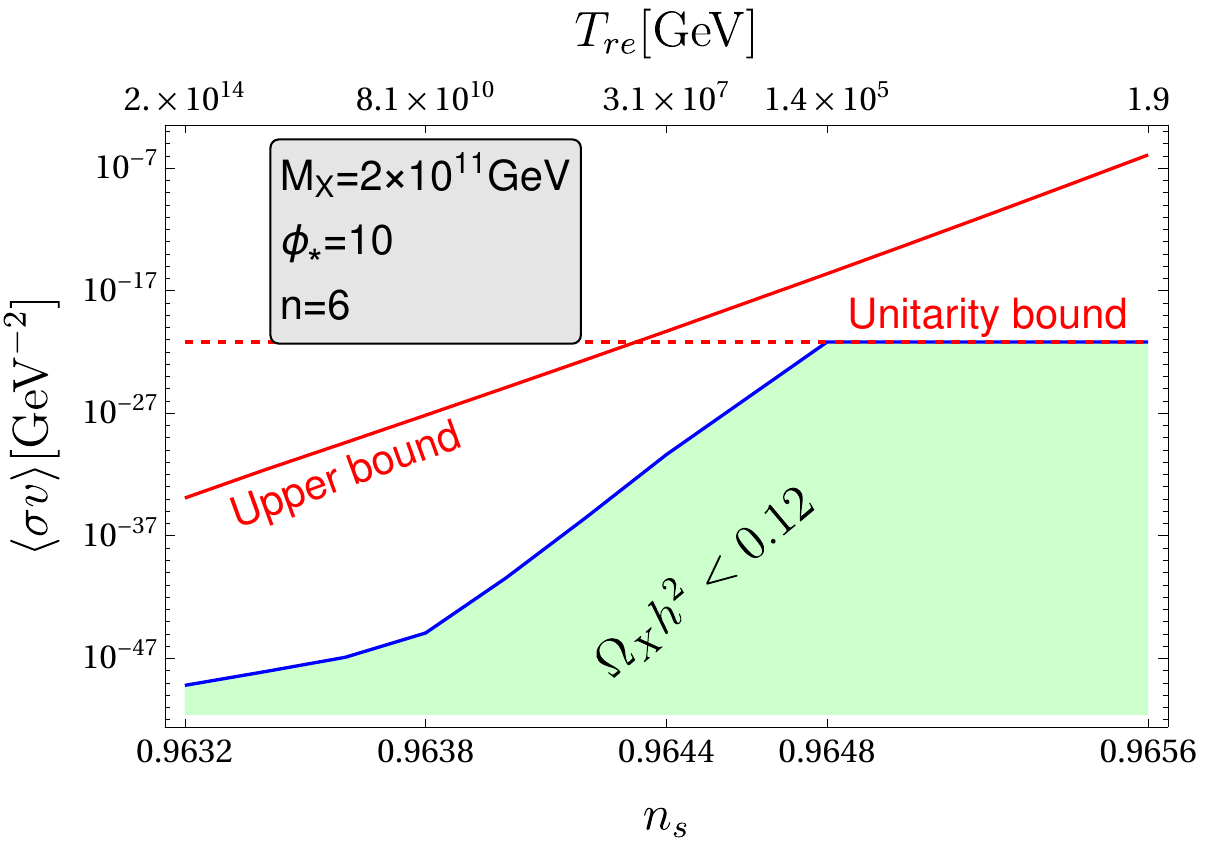}}
	\caption{$(\langle\sigma v\rangle~vs~n_s)$ were plotted for three values of $\phi_*$ and $n=6$. For all the cases we have chosen a dark matter mass $M_X=2\times 10^{11}  GeV$.}
	\label{n6nssigma}
\end{figure}
\begin{figure}[t]
	\centering	
	\subfigure[]{\includegraphics[scale=0.3]{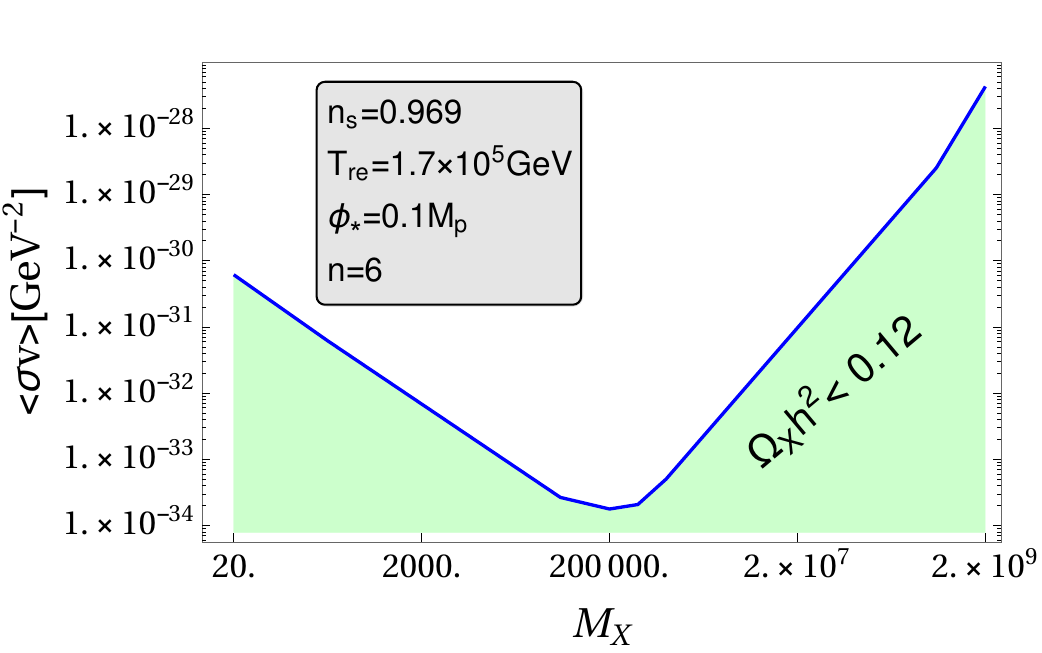}}
	\subfigure[]{\includegraphics[scale=0.3]{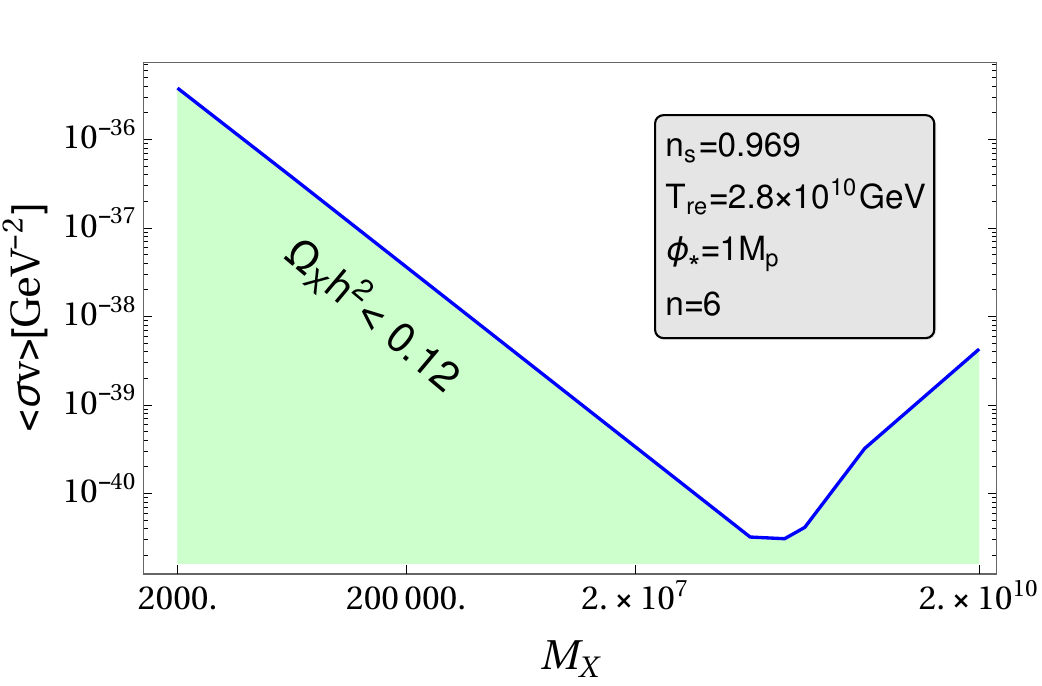}}
	\subfigure[]{\includegraphics[scale=0.3]{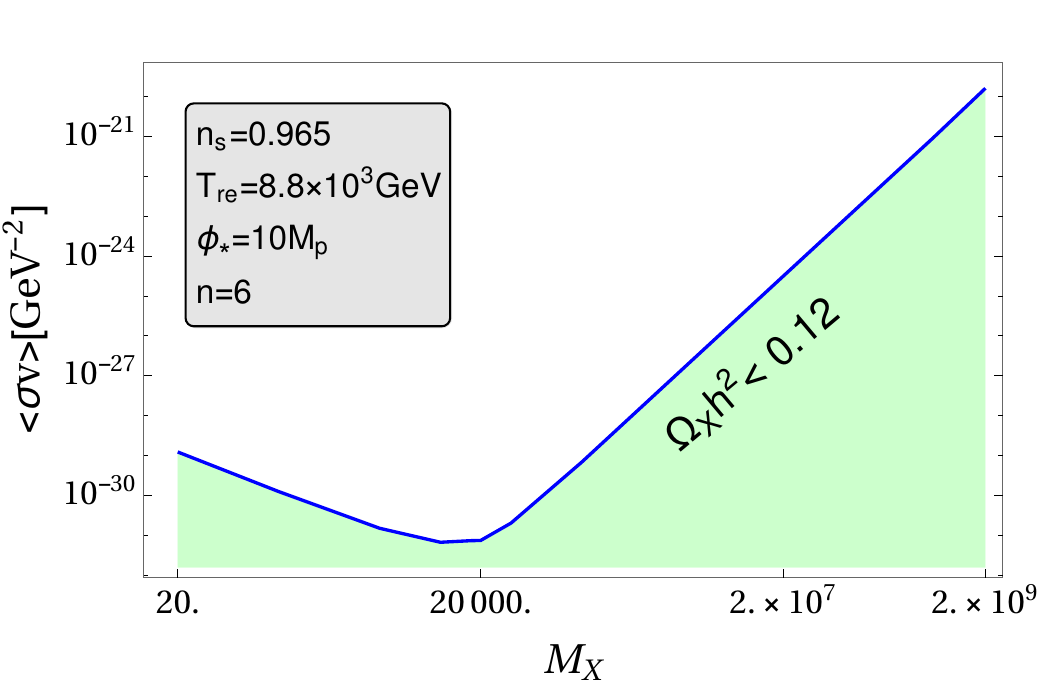}}
	\caption{$(\langle\sigma v\rangle~vs~M_X)$ were plotted for three values of $\phi_*$ and $n=6$. General descriptions of the plots are same as before.}
	\label{n6mxsigma}
\end{figure} 
\begin{table}[h!]
	\label{modeln6}
	Model parameters and associated constraints on dark matter parameters for n=6
	\begin{minipage}{.33\linewidth}
		\caption{$\phi_*=0.1M_p$}
		\begin{tabular}{|c|c|c|c}\hline
			{$n_s$} & {$N_{re}$} & {$T_{re}(GeV)$} & $\begin{array}{c}{\fontsize{7}{7}\selectfont \textit{$\langle \sigma v\rangle GeV^{-2}$}}\\ {\fontsize{7}{7}\selectfont \textit{$M_X(10^{11} GeV)$}}\end{array}$\\\hline
			{0.9678} & {3} & ${4\times 10^{13}}$ & ${3\times 10^{-49}}$\\
			{0.968} & {5} & ${2 \times 10^{12}}$ & $7\times 10^{-48}$\\
			{0.969} & {20} & $2\times 10^5$ & $9\times 10^{-34}$\\
			{0.9698} & {32} & {0.2} & $6\times 10^{-22}$\\\hline
		\end{tabular}
	\end{minipage}%
	\begin{minipage}{.33\linewidth}
		\caption{$\phi_*=1M_p$}
		\begin{tabular}{|c|c|c|c}\hline
			\textit{$n_s$} & \textit{$N_{re}$} & \textit{$T_{re}(GeV)$} & $\begin{array}{c}{\fontsize{7}{7}\selectfont \textit{$\langle \sigma v\rangle GeV^{-2}$}}\\ {\fontsize{7}{7}\selectfont \textit{$M_X(10^{11} GeV)$}}\end{array}$\\\hline
			{0.9682} & {2} & ${1\times 10^{14}}$ & ${9\times 10^{-50}}$\\
			{0.969} & {14} & ${3\times 10^8}$ & ${1\times 10^{-36}}$\\
			{0.970} & {29} & {10.58} & $unitarity$\\
			{0.972} & {32} & {0.3} & $unitairty$\\\hline
		\end{tabular}
	\end{minipage}%
	\begin{minipage}{.32\linewidth}
		\caption{$\phi_*=10M_p$}
		\begin{tabular}{|c|c|c|c|}\hline
			\textit{$n_s$} & \textit{$N_{re}$} & \textit{$T_{re}(GeV)$} & $\begin{array}{c}{\fontsize{7}{7}\selectfont \textit{$\langle \sigma v\rangle GeV^{-2}$}}\\ {\fontsize{7}{7}\selectfont \textit{$M_X(10^{11} GeV)$}}\end{array}$\\\hline
			{0.9632} & {2} & ${1.9\times 10^{14}}$ & $6\times10^{-50}$\\
			{0.964} & {11} & ${6\times 10^9}$ & ${4\times 10^{-41}}$\\
			{0.965} & {23} & ${8.8\times 10^3}$ & ${unitarity}$\\
			{0.9656} & {30} & {2} & ${unitarity}$\\\hline
		\end{tabular}
	\end{minipage}%
\end{table}
In this sub-section we summarized all our analysis in the figs.(\ref{n6_nsnret},\ref{n6nssigma},\ref{n6mxsigma}) and the following tables (\ref{modeln6}) for $n=6$.

 Important difference with $n<4$ models turned out to be the slope of $(n_s~ vs~T_{re})$ and $(n_s~vs~N_{re})$ curve. This has also been explained in the schematic diagram fig.(\ref{scales}). This is coming from the stiff equation of state of the oscillating inflaton $w_{\phi} = (n-2)/(n+2) = 1/2 > 1/3$. The duration of reheating period ($N_{re}$) increases with increasing $n_s$. Therefore, unlike $n<4$ models, corresponding to the maximum reheating temperature we have minimum possible values of $n_s$. 
\subsubsection{Results and constraints: {\bf Model n=8}}

For $n=8$, the qualitative behavior of all the plots will be same as $n=6$. Therefore we summarize all our results in the 
table-\ref{modeln8} and the respective plots. 
Important to see that very small reheating temperature can be reached in this model. At the central value of $n_s=0.968 \pm 0.006$, we see reheating temperature could be $10^4\sim 10^5$ GeV given in the tab.(\ref{modeln8}). The lowest value of $n_s$ can be set from the minimum possible reheating temperature coming from BBN constraints\cite{Kawasaki:1999na,Kawasaki:2000en,Steigman:2007xt,Fields:2014uja}, which is $T_{re} \sim 0.1GeV$.
\begin{figure}[t!]
	\centering	
	\subfigure[]{\includegraphics[scale=0.45]{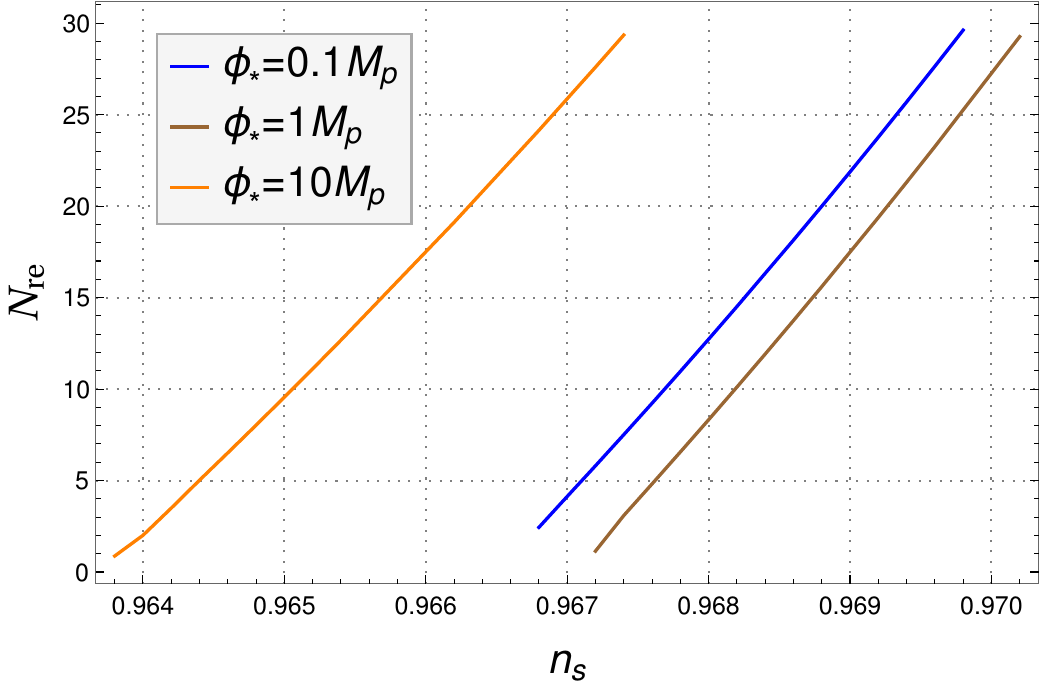}}
	\subfigure[]{\includegraphics[scale=0.45]{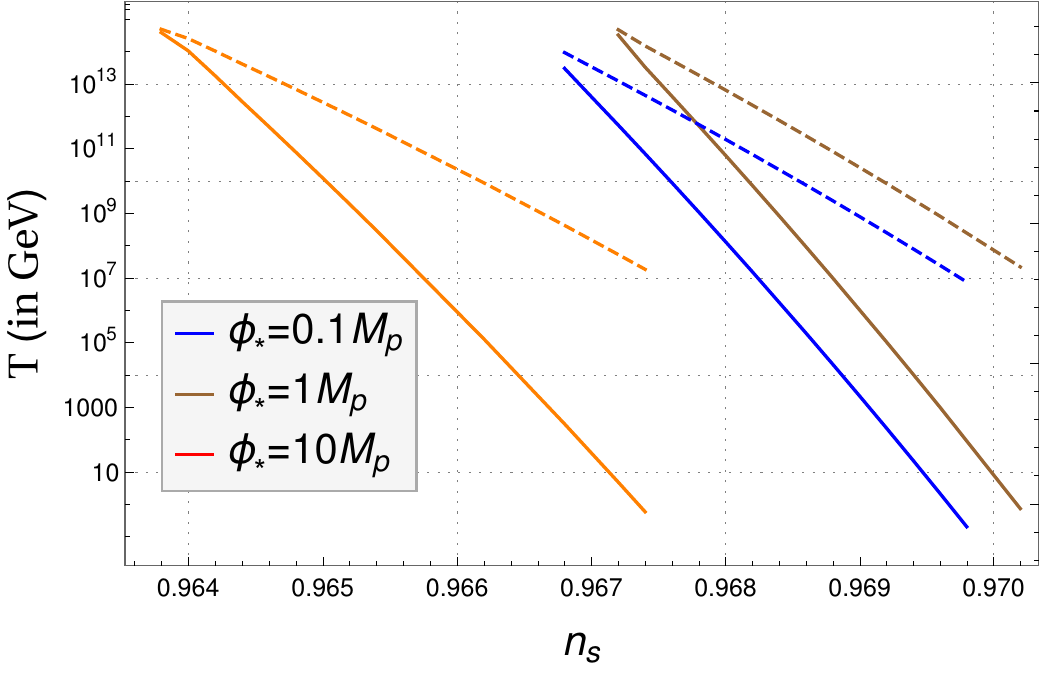}}
	\caption{Evolution of (a) Reheating e-folding number $N_{re}$ during reheating and (b) the reheating temperature $T_{re}$(solid lines) and the maximum temperature $T_{max}$(dotted lines) with respect to $n_s$ for $n=8$ with the same behavior as $n=6$.}
	\label{n8_nsnret}
\end{figure}
\begin{table}[h!]
	\label{modeln8}
	Model parameters and associated constraints on dark matter parameters for n=8
	\begin{minipage}{.33\linewidth}
		\caption{$\phi_*=0.1M_p$}
		\begin{tabular}{|c|c|c|c}\hline
			{$n_s$} & {$N_{re}$} & {$T_{re}(GeV)$} & $\begin{array}{c}{\fontsize{7}{7}\selectfont \textit{$\langle \sigma v\rangle GeV^{-2}$}}\\ {\fontsize{7}{7}\selectfont \textit{$M_X(10^{11} GeV)$}}\end{array}$\\\hline
			{0.9668} & {2} & ${3 \times 10^{13}}$ & ${4 \times 10^{-49}}$\\
			{0.968} & {13} & ${1 \times 10^8}$ & ${7 \times 10^{-35}}$\\
			{0.969} & {22} & ${2 \times 10^{3}}$ & \textit{unitairty}\\
			{0.9698} & {30} & {0.2} & \textit{unitarity}\\\hline
		\end{tabular}
	\end{minipage}%
	\begin{minipage}{.33\linewidth}
		\caption{$\phi_*=1M_p$}
		\begin{tabular}{|c|c|c|c}\hline
			\textit{$n_s$} & \textit{$N_{re}$} & \textit{$T_{re}(GeV)$} & $\begin{array}{c}{\fontsize{7}{7}\selectfont \textit{$\langle \sigma v\rangle GeV^{-2}$}}\\ {\fontsize{7}{7}\selectfont \textit{$M_X(10^{11} GeV)$}}\end{array}$\\\hline
			{0.9672} & {1} & ${3 \times 10^{14}}$ & ${4 \times 10^{-50}}$\\
			{0.968} & {8} & ${6 \times 10^{10}}$ & ${2 \times 10^{-45}}$\\
			{0.9692} &{19} & ${1 \times 10^5}$ & \textit{unitarity}\\
			{0.9702} & {29} &{0.72} & \textit{unitarity}\\\hline
		\end{tabular}
	\end{minipage}%
	\begin{minipage}{.32\linewidth}
		\caption{$\phi_*=10M_p$}
		\begin{tabular}{|c|c|c|c|}\hline
			\textit{$n_s$} & \textit{$N_{re}$} & \textit{$T_{re}(GeV)$} & $\begin{array}{c}{\fontsize{7}{7}\selectfont \textit{$\langle \sigma v\rangle GeV^{-2}$}}\\ {\fontsize{7}{7}\selectfont \textit{$M_X(10^{11} GeV)$}}\end{array}$\\\hline
			{0.9638} & {1} & ${4\times 10^{14}}$ & ${3 \times10^{-50}}$\\
			{0.964} & {2} & ${1\times 10^{14}}$ &${1 \times 10^{-49}}$\\
			{0.965} & {10} & ${1 \times 10^{10}}$ & ${3 \times 10^{-43}}$\\
			{0.9674} & {29} & {1} & \textit{unitarity}\\\hline
		\end{tabular}
	\end{minipage}%
\end{table}

\begin{figure}[h!]
	\centering	
	\subfigure[]{\includegraphics[scale=0.3]{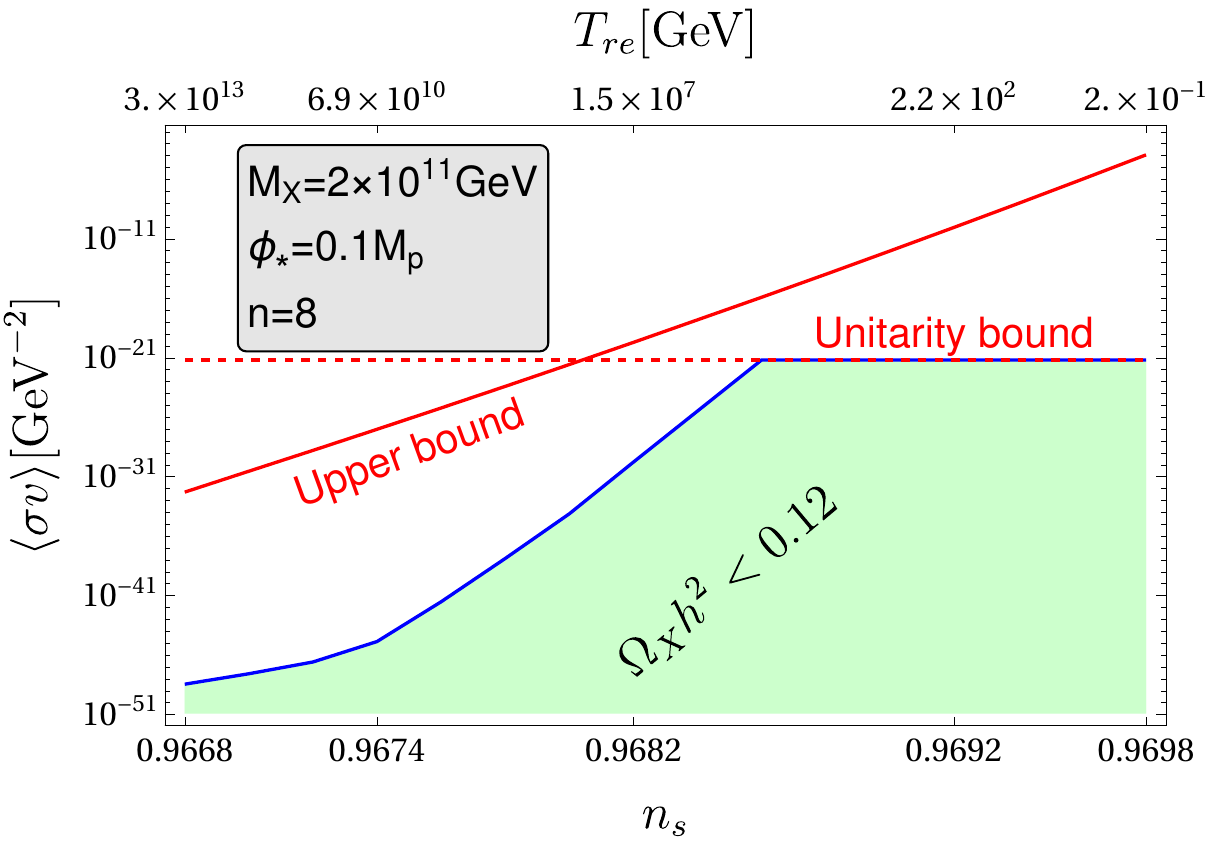}}
	\subfigure[]{\includegraphics[scale=0.3]{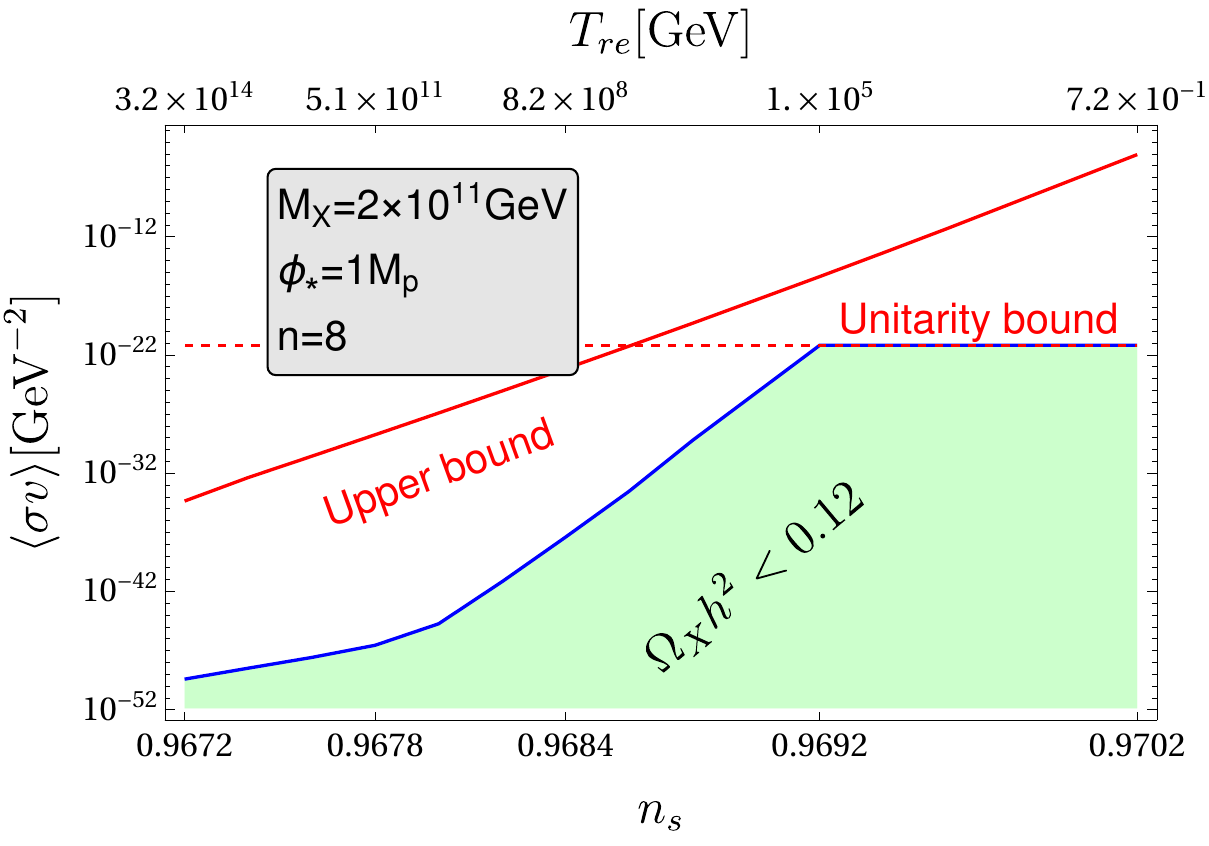}}
	\subfigure[]{\includegraphics[scale=0.3]{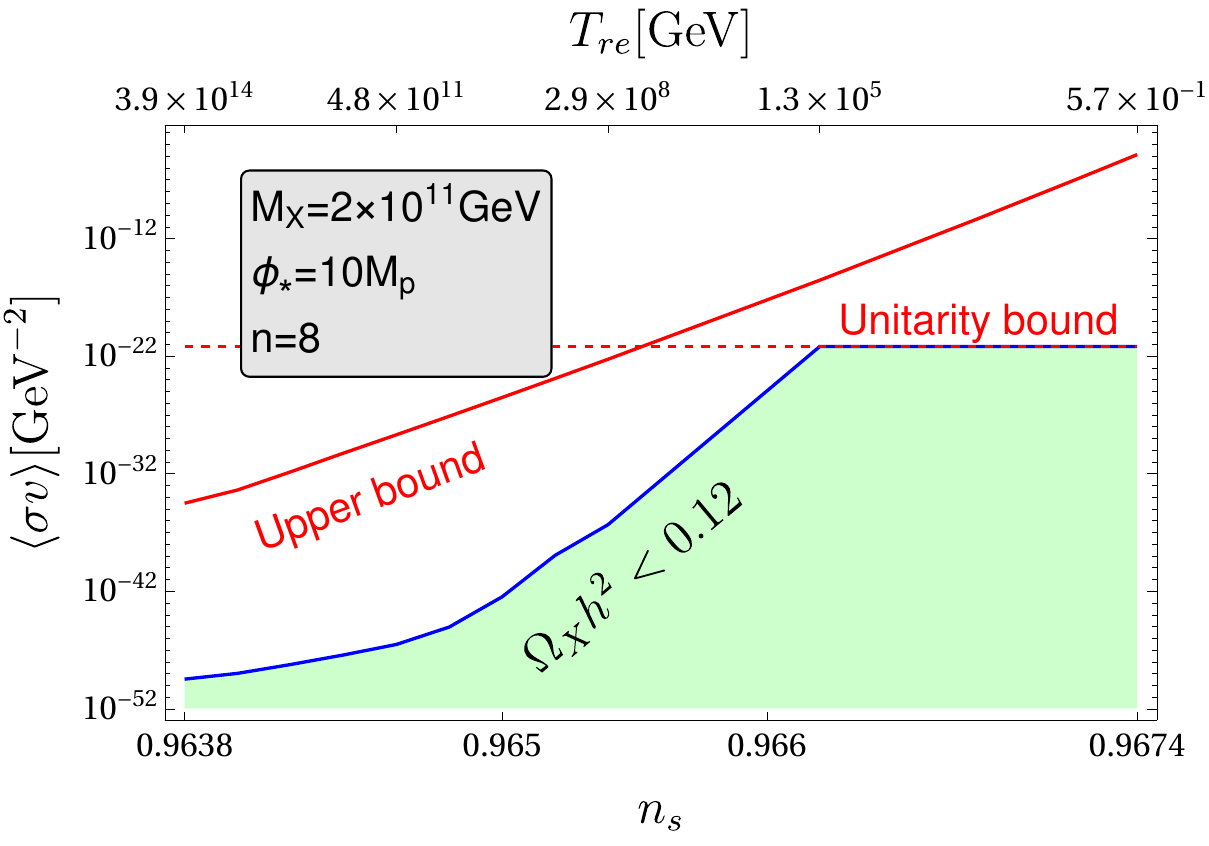}}
	\caption{$(\langle\sigma v\rangle~vs~n_s)$ were plotted for $n=8$. Descriptions are same as  previous plots. For all the plots we have chosen dark matter mass $M_X=2\times 10^{11} GeV$.}
	\label{n8nssigma}
\end{figure}

\begin{figure}[t]
	\centering	
	\subfigure[]{\includegraphics[scale=0.3]{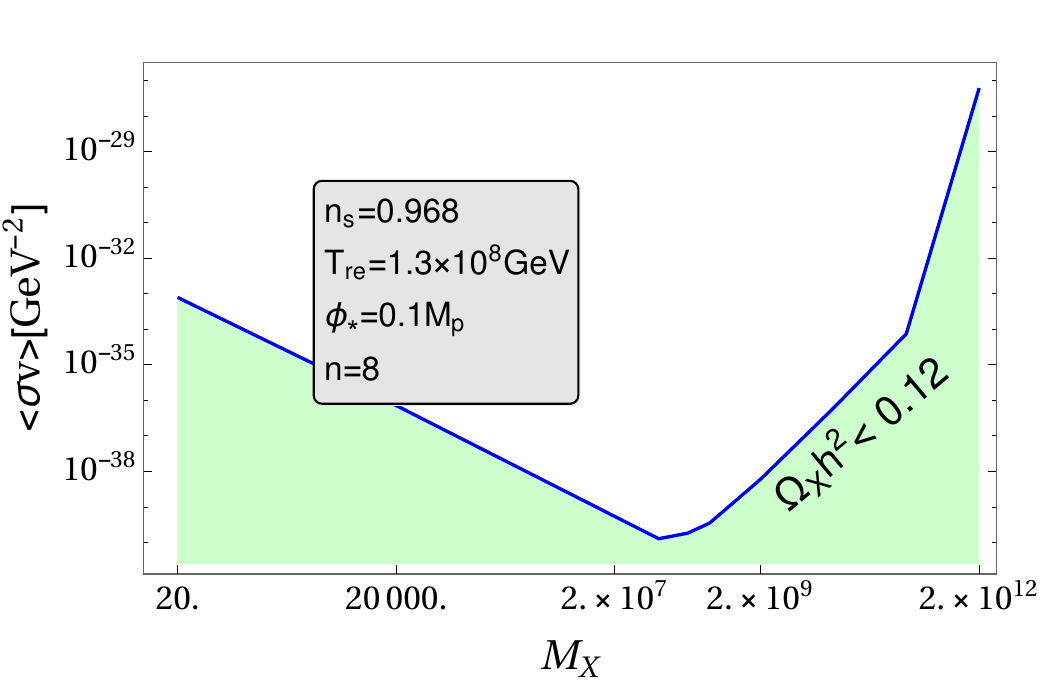}}
	\subfigure[]{\includegraphics[scale=0.3]{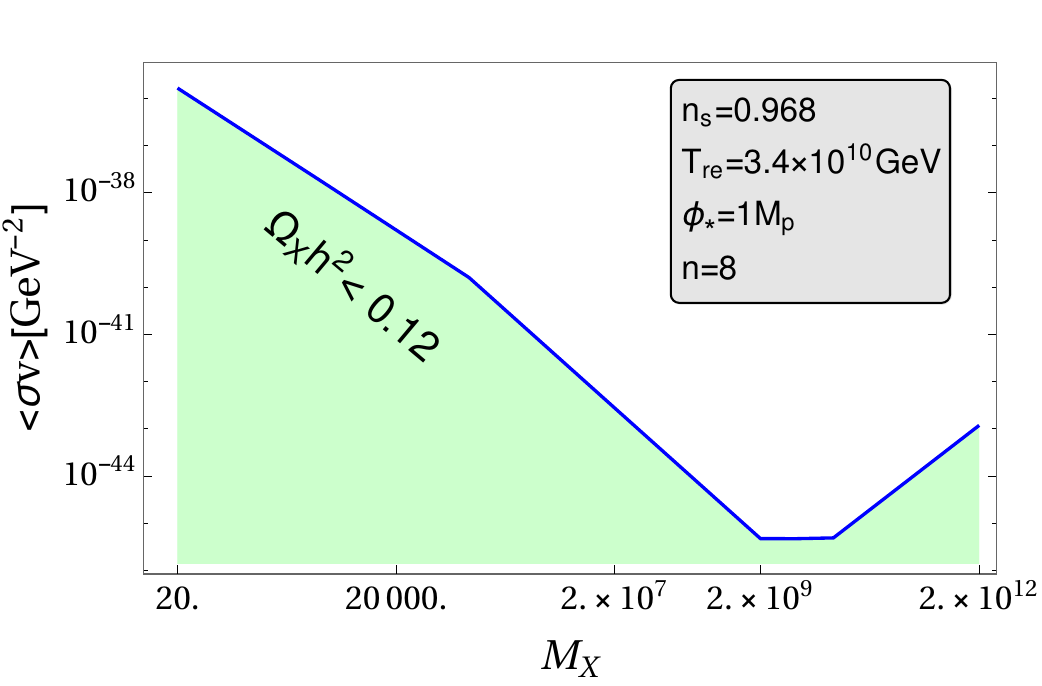}}
	\subfigure[]{\includegraphics[scale=0.3]{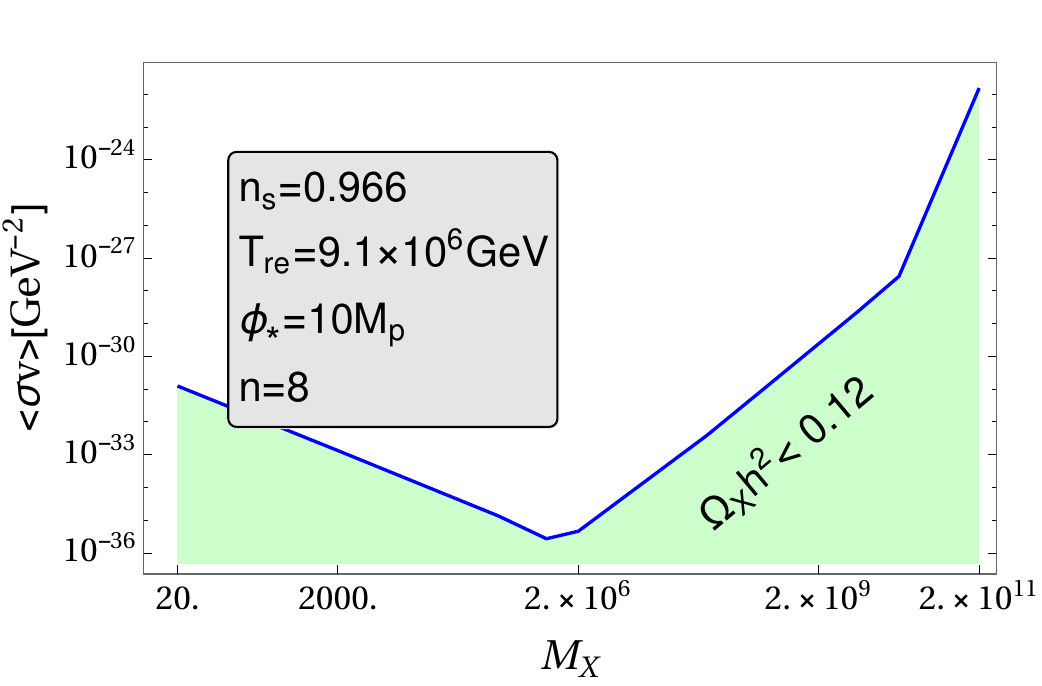}}
	\caption{$(\langle\sigma v\rangle~vs~M_X)$ were plotted for $n=8$.}
	\label{n8_Mxsigma}
\end{figure}

\newpage
\section{\label{conclusion} Conclusions and future directions}
In this paper, we have done the reheating constraint analysis in detail for a class of inflationary models with non-standard potential \cite{Maity:2019ltu}. Our focus was on understanding the inflation model as well as dark matter phenomenology via reheating phase. For both the analysis, the CMB plays a very important role. We, therefore, follow the formalism developed in \cite{Maity:2018dgy} where the dark matter parameter space is constrained via CMB anisotropy and the reheating dynamics. Constraints on dark matter parameter space are dependent upon the specific model of inflation. As mentioned throughout the paper, in our recent work on minimal inflationary cosmologies we had discussed in detail the role of inflationary scale $\phi_*$. By tuning it, we also obtained both high scale as well as low scale inflations which are observationally viable. In this paper we considered those inflationary models and further extended our work on understanding the role of that scale on the reheating phase as well as on dark matter phenomenology.

As we all know, the reheating phase, which is relatively less understood due to lack of direct observables, has potential to explain many unanswered questions related to dark matter, baryogenesis etc. In this paper, we have studied in detail the reheating period considering the perturbative decay of inflaton into radiation and radiation into dark matter component.
One of the important findings from our analysis is the existence of maximum reheating temperature $T_{re}^{max}\simeq 10^{15}$ GeV, which is independent of large or small scale inflation model. Therefore, it is the CMB scale which dictates the value of maximum possible reheating temperature. Corresponding to that maximum $T_{re}$, we also have either maximum possible $n_s^{max}$ for all the models with $n<4$ or minimum possible $n_s^{min}$ for models with $n>4$. This fact can be easily understood from the illustration in fig.\ref{scales}.    
As we have already emphasized the goal was to connect the inflation, reheating phase with the CMB anisotropy and the dark matter abundance. Therefore, we further study how the inflationary scalar spectral index fixes the dark matter annihilation cross-section for a given mass of the dark matter once we specify the model. Here is our main findings which tells us how uniquely  we can constrain the dark matter parameter space $(M_X, \langle \sigma v\rangle)$, through the following important relations,   
	\bea \label{abundance1}
	&&\Omega_X h^2 \propto  \left<\sigma|v|\right> M_X^4 Exp\left[-\frac{E(w_{\phi})M_X}{T_{max}}\right]~~ \text{for $M_X \gtrsim T_{max}$} \nno\\
	&&\Omega_X h^2  \propto  \frac{\left< \sigma v\right>}{M_X^{\frac{9-7\wI}{2(1+\wI)}}} \left[\left(\frac{a_0 T_0}{k} \right) H_k e^{-N_k} e^{-N_{re}}\right]^{\frac{7+3\wI}{1+\wI}}\qquad \text{for~} T_{max} > M_X > T_{re} \nno\\
	&&\Omega_X h^2\propto  \left< \sigma v\right> M_X \left(\frac{a_0 T_0}{k} \right) H_k e^{-N_k} e^{-N_{re}} \qquad \text{When~}M_X < T_{re}.
	\eea 
The non-trivial relation among the dark matter abundance $(\Omega_X h^2)$ , the general inflaton equation state $(\omega_{\phi})$ , CMB scale $k$, and mass of the dark matter $M_X$ is our main result which has not been reported before. 
Here an important point we would like to emphasize again that the dark matter is produced by Freeze-in mechanism. Therefore, detailed analysis for freeze out dark matter based on our formalism will be reported elsewhere.

Nevertheless
considering the central value of $n_s \sim 0.968$ as given by PLANCK our model predictions are summarized in table-\ref{summarycon}.
\begin{table}[t]
	\label{summarycon}
Summary table for central value of $n_s$ with dark matter mass taken to be $M_X =200$ TeV
	\begin{minipage}{.32\linewidth}
		\caption{$\phi_*=0.1M_p$}
		\begin{tabular}{|c|c|c|c|}\hline
			{$n$} & {$N_{re}$} & {$T_{re}(GeV)$} & $\begin{array}{c}{\fontsize{7}{7}\selectfont \textit{$\langle \sigma v\rangle GeV^{-2}$}}\\ {\fontsize{7}{7}\selectfont \textit{$M_X(200 TeV)$}}\end{array}$\\\hline
			{2} & {33} & {$2.4\times 10^{4}$}   & {$3\times 10^{-30}$}\\
			{4} & {-}  & {-} &  {-}\\
			{6} & {5} & ${1.6 \times 10^{12}}$ & $7\times 10^{-42}$\\
			{8} & {-} & {-} & {-}\\
			\hline
		\end{tabular}
	\end{minipage}%
	\begin{minipage}{.32\linewidth}
		\caption{$\phi_*=1M_p$}
		\begin{tabular}{|c|c|c|c|}\hline
			{$n_s$} & {$N_{re}$} & {$T_{re}(GeV)$} & $\begin{array}{c}{\fontsize{7}{7}\selectfont \textit{$\langle \sigma v\rangle GeV^{-2}$}}\\ {\fontsize{7}{7}\selectfont \textit{$M_X(200 TeV)$}}\end{array}$\\\hline
			{2}  & {31}   & {$1.4\times 10^5$}   & {$6\times 10^{-34}$}\\
			{4} & {-}  & {-} &  {-}\\
			{6} & {13} & ${1.3\times 10^{8}}$ & ${7\times 10^{-38}}$\\
			{8} & {8} & ${6 \times 10^{10}}$ & ${2 \times 10^{-40}}$\\
			\hline
		\end{tabular}
	\end{minipage}%
	\begin{minipage}{.32\linewidth}
		\caption{$\phi_*=10M_p$}
		\begin{tabular}{|c|c|c|c|}\hline
			\textit{$n$} & \textit{$N_{re}$} & \textit{$T_{re}(GeV)$} & $\begin{array}{c}{\fontsize{7}{7}\selectfont \textit{$\langle \sigma v\rangle GeV^{-2}$}}\\ {\fontsize{7}{7}\selectfont \textit{$M_X(200 TeV)$}}\end{array}$\\\hline
			{2} & {1.5} & ${6\times10^{14}}$  & {$3\times10^{-44}$}\\
			{4} & {-}  & {-} &  {-}\\
			{6} & {-}  & {-} &  {-}\\
			{8} & {-}  & {-} &  {-}\\
			\hline
		\end{tabular}
	\end{minipage}%
\end{table}
For $n=4$, as we have already mentioned the data is unconstrained and is plagued by numerical artifact. Therefore, we left those parameters blank in the summary table. From the above summary table we can clearly see that given the central value of $n_s = 0.968$ and dark matter mass $M_X =200$ TeV, few models can be ruled out based on the values of the model parameters for which prediction will be out of $1\sigma$ range of $n_s$ given by PLANCK. On the other hand, if one considers $n_s$ within the given range of PLANCK, we can give a bound on the possible range of values of annihilation cross-section for a given mass. An Important conclusion that we can make is that once the CMB anisotropy measurement fixes the value of $n_s$, we can completely fix the dark matter annihilation cross-section for a given mass considering the perturbative reheating process.  
Therefore, it is very important to pinpoint the actual value of $n_s$ in the future experiment.
Understanding the dark matter phenomenology is one of the important challenges in the particle physics community. Therefore, our formalism and present analysis may shed some light on this subject. The important assumption of our analysis is the perturbative reheating. It is well known that the non-perturbative effects such as parametric resonance can be very efficient that can initiate the radiation dominance within few efolding numbers after the end of inflation. {We have studied the effects of non-perturbative particle production for the plateau inflation model in\cite{Maity:2018qhi}. It has been found that for inflation potential with $n>2$, the equation of the state of the system will behave as radiation within a few e-folding number after the initiation of preheating irrespective of the values of the scale $\phi_{\ast}$. However for $n=2$ potential, the equation of state do not became that of the radiation with the type of interaction term considered in the work. As a result, for $n=2$ case, we have done the perturbative analysis taking the system after the preheating as initial state. The general conclusion for such a two stage reheating is that the minimum $n_s$ further increases that constrain the range of allowed $n_s$ more. However, for $n>2$ cases the initial stage after preheating will be similar to that of the $n=4$ case, in which case the CMB will have very little information regarding the reheating phase.} However, as shown in\cite{GarciaBellido:2008ab}, incorporating interactions among the produced particles, the parametric resonance can be delayed and as a result we will get an extended period of reheating. The validity of the perturbative regime in the context of reheating has been considered in\cite{Drewes:2017fmn} considering the effective equation of state. However, the issue of self-resonance still remains. It has been shown in\cite{Lozanov:2016hid,Lozanov:2017hjm} that when considering the self-resonance, the homogeneous condensate will cease oscillating after a certain period, and ends up fragmenting into higher momentum modes of the inflaton field. For, certain form of the potentials, $V(\phi)\propto \phi^n$, the equation of state eventually ends up becoming $1/3$ instead of $(n-2)/(n+2)$. However, there is a certain period when this equation of state will be maintained and moreover, there are specific parameter range when the self-resonance will be inefficient. We are currently working on these issues of taking into account the non-perturbative effects.
Finally at the end let us point out an another important direction which we will report elsewhere. In our study we have solved for the homogeneous Boltzmann equation for all the energy components. However, it is well known that evolution of perturbation for every individual component would be extremely important to understand the small scale properties of CMB. Interestingly in our unified approach inflation and reheating dynamics control all the evolution. Therefore, any small scale CMB observables will certainly contain valuable information related to inflation. In this regard well known small scale $\mu$-type and $y$-type spectral distortions of CMB are extremely important. At present those distortion parameters are tightly constrained by COBE and FIRAS experiments, $|\mu| < 9 \times 10^{−5}$ and $ y < 1.5 \times 10^{−5}$ \cite{cobe}. However, future projected sensitivity for PIXIE \cite{pixie} and PRISM \cite{prism} experiments are within $ 10^{-8}\sim 10^{-9}$.The standard model of particle physics interactions already predicts those spectral distortion parameters to be $\sim 10^{-8}$ \cite{spectral1,spectral2}. Keeping in mind the future experimental sensitivity, it would be extremely important to understand various other physics processes which can give rise to those distortions. Therefore, by using our unified approach we can further constrain inflation models considering those distortion parameters.

\section{Acknowledgement}
We thank the HEP and Gravity group members at the department for their comments and discussions. We thank the referee for valuable comments which helped in improving the work.

\hspace{0.5cm}

\end{document}